\documentclass[aip,jcp,amsmath,amssymb,floatfix,reprint,noeprint,superscriptaddress,twocolumn,citeautoscript]{revtex4-2}

\usepackage[version=3]{mhchem} 
\usepackage{listings}
\usepackage{amsmath}
\usepackage{amssymb}
\usepackage{braket}
\usepackage{wrapfig}
\usepackage{graphicx}
\usepackage{dcolumn}
\usepackage{bm}
\usepackage{siunitx}
\DeclareSIUnit\angstrom{\text {Å}}
\usepackage[colorlinks=true,linkcolor=blue,citecolor=blue,urlcolor=blue]{hyperref}
\usepackage{tabularx}

\begin{document}

\preprint{AIP/123-QED}

\title[]{Benchmarking short-range machine learning potentials for atomistic simulations of metal/electrolyte interfaces}

\author{Lucas B.T. de Kam}
\affiliation{Leiden Institute of Chemistry, Leiden University, 2300 RA Leiden, the Netherlands}
\author{Jia-Xin Zhu}
\affiliation{State Key Laboratory of Physical Chemistry of Solid Surfaces, iChEM, College of Chemistry and Chemical Engineering, Xiamen University, Xiamen 361005, China}
\author{Ankit Mathanker}
\affiliation{Department of Chemical Engineering, University of Michigan, Ann Arbor, MI, USA}
\author{Katharina Doblhoff-Dier}
\email{k.doblhoff-dier@lic.leidenuniv.nl}
\affiliation{Leiden Institute of Chemistry, Leiden University, 2300 RA Leiden, the Netherlands}

\author{Nitish Govindarajan}
\email{nitish.govindarajan@ntu.edu.sg}
\affiliation{School of Chemistry, Chemical Engineering and Biotechnology, Nanyang Technological University, Singapore 637371, Singapore}

\date{\today}

\begin{abstract}
Atomistic simulations of electrochemical interfaces remain challenging due to the long time scales required to adequately sample the structure of the electric double layer. The emergence of efficient, short-range machine learning interatomic potentials (MLIPs) offers a promising alternative to computationally expensive density functional theory-based molecular dynamics (DFT-MD) simulations in this regard. However, in standard periodic DFT calculations of metal surfaces, the surface charge is implicitly set by the number of counterions in the simulation cell, making it a global property that is difficult to represent with strictly local MLIPs. Here, we benchmark common MLIP architectures (DP, ACE, MACE) for charged Au/water interfaces containing solvated sodium ions. We find that MLIPs trained on datasets spanning multiple surface charge states yield inconsistent predictions of interfacial water orientation and ion distributions, although message-passing models with a larger receptive field exhibit greater robustness to training on mixed-charge datasets. In contrast, models trained on a single charge state produce consistent equilibrium interfacial properties. Finally, we assess the performance of the eSEN model trained on the recently released Open Catalyst 2025 dataset, which includes solid/liquid interfaces that span a wide range of surface charge densities. Overall, our results characterize the limitations of short-range MLIPs for simulations of electrochemical interfaces and provide practical guidance for constructing training datasets for simulations of charged metal/electrolyte interfaces.
\end{abstract}

\maketitle


\section{Introduction}
Electrochemical reactions take place within the electric double layer, formed by the accumulation of (counter)charged electrolyte ions and polarized solvent molecules at a charged electrode surface. Molecular dynamics (MD) simulations are widely used to investigate how electrode charge and electrolyte composition shape the structure of the double layer \citep{alfarano2021stripping, tran2024hydration, moss2025if, becker2023multiscale}, and how this structure affects kinetics \citep{qin2024modulating, shah2024platinum, li2025kinetic, tian2025electrochemical}. Metal–water interfaces commonly serve as a model system. Classical force fields enable MD simulations of aqueous double layers, which typically extend over a few nanometers, for hundreds of nanoseconds \citep{tran2024hydration}. However, they rely on empirical formulations and do not treat electrons explicitly. Consequently, they struggle to capture electronic effects in the double layer (e.g. bond breaking, charge transfer, and electron spillover) and require careful parametrization.

Quantum mechanics (QM) methods can accurately describe electronic effects. In particular, density functional theory-based MD (DFT-MD) simulations have played a pivotal role in shaping the modern understanding of metal/water interfaces \citep{le2017determining, le2018structure, le2018theoretical, le2020molecular, gross2022ab, dominguez2024metal}. However, these simulations are typically limited to system sizes of $<10^3$ atoms and a timescale of $< \SI{e2}{\pico\second}$. With a typical diffusion coefficient of $D\approx\SI{e-5}{\centi\meter\squared\per\second}$ for solvated ions \citep{Haynes2016CRC}, a simulation time of at least $t\sim\langle x^2\rangle/2D=\SI{0.5}{\nano\second}$ is required to equilibrate solvated ions over a length of \SI{1}{\nano\meter}. This timescale is beyond what is usually achievable with DFT-MD, especially for a systematic study of different systems. Simulating charged interfaces with DFT-MD therefore often involves harsh approximations, such as relying on short trajectories, replacing part of the electrolyte with a continuum model, or using poorly converged DFT. The effect of these approximations is often unknown and difficult to estimate \citep{govindarajan2025intricacies}.

Recent developments in the field of machine learning have made it possible to fit machine learning interatomic potentials (MLIPs) to energies and forces from QM calculations \citep{thiemann2024introduction}. Often, fewer than a couple of thousand QM calculations are required to train accurate application-specific MLIPs \citep{cheng2019ab, stark2024benchmarking, zhu2025dielectric, guo2025understanding}. These surrogate models allow atomistic simulations to be performed on length and time scales far beyond those achievable by DFT-MD, extending the scope of simulations by orders of magnitude \citep{morrow2022indirect, guo2025understanding, zhu2025dielectric}. Therefore, MLIPs promise improved confidence and reproducibility of simulation results. The reduced number of QM calculations enables the use of tighter convergence settings or more accurate methods (beyond DFT) \citep{daru2022coupled, o2024pair}. Larger timescales allow for improved sampling statistics and careful equilibration of thermostats. 

However, the choice of MLIP architecture and model training introduces additional complexity in performing molecular dynamics simulations compared to DFT-MD. Given the rapid development in MLIPs, it is often unclear how these choices affect the simulation outcome. Benchmarking studies of MLIPs offer valuable insight into these challenges. For example, Refs.~\onlinecite{stark2024benchmarking} and \onlinecite{leimeroth2025machine} benchmarked various MLIPs for gas-surface dynamics and complex solids, respectively. These studies found that while graph neural network architectures are accurate, they are computationally expensive, whereas simpler models, such as atomic cluster expansion (ACE) potentials, are much faster but not always sufficiently accurate.

Charged interfaces possess unique properties that are absent in the aforementioned benchmarking studies. An important issue is the need for long-range electrostatic interactions. Most MLIPs are based on the \textit{locality assumption}, which posits that the properties of an atom depend only on its local environment \citep{behler2021machine, yue2021short}. The application of MLIPs to large interfacial systems therefore requires careful treatment of electrostatics, which remains an active area of research \citep{zhu2025machine, wang2025ion, yang2025macro}. However, in electrocatalysis, simulations often focus on local chemical interactions between reactants and electrolyte species in the Helmholtz layer --- typically within \SI{1}{\nano\meter} from the surface \citep{govindarajan2025intricacies, li2025kinetic, qin2024modulating}. Under the assumption of a high ion concentration and correspondingly short screening lengths, short-range MLIPs may capture the majority of relevant interactions. Although this assumption enables less expensive simulations by avoiding explicit long-range electrostatics, its validity remains poorly understood. Understanding how well short-range MLIPs can describe the interfacial region is therefore the first objective of this study.

Ideally, a single MLIP could be trained to describe multiple surface charge states simultaneously. Using a single model across different charge states increases transferability, which in turn greatly reduces human effort and computational cost associated with training and dataset construction. In practice, however, this task is far from trivial. Standard periodic DFT calculations require the simulation supercell to maintain overall charge neutrality. To model a charged interface under these constraints, one common strategy is to introduce explicit counterions near the metal surface \citep{le2020molecular, govindarajan2025intricacies}. For example, species such as sodium (\ce{Na}), whose valence energy level lies above the metal Fermi level, typically transfer their valence electron to the metal. This results in a negatively charged metal surface and a corresponding positively charged counterion in the electrolyte region. Notably, the net surface charge in such simulations is not a strictly local quantity but rather a global property determined by the total number of counterions in the supercell. MLIPs trained on DFT data for explicitly charged interfaces must explicitly `see' the ions in order to learn the correct electrostatic and structural response of the interface. The second objective of this work is therefore to benchmark short-range MLIPs with respect to their ability to learn and generalize interfacial behavior across different imposed surface charges. We then compare their performance to that of an MLIP trained on the Open Catalyst 2025 (OC25) dataset \citep{sahoo2025open}, which contains approximately 8 million solid/liquid interface configurations spanning a wide range of surface charges.

In our benchmark, we include the Deep Potential (DP) \citep{zhang2018end} and its message-passing variant (DP-MP) \citep{gao2024enhanced}, ACE \citep{bochkarev2024graph, drautz2019atomic}, and MACE \citep{batatia2022mace, batatia2025design}. The main properties we use for evaluation are the interfacial water orientation and spatial distribution of ions at the interface, both of which are highly sensitive to the imposed surface charge \citep{alfarano2021stripping, guo2025understanding, le2020molecular}. By systematically comparing the performance of short-range MLIPs for charged metal-electrolyte interfaces, we identify when these models can reliably reproduce interfacial properties. This enables users to make informed decisions about when MLIP-based acceleration is appropriate, and when caution is needed. We also point out current model limitations and suggest directions for developing more robust and transferable MLIPs for simulations of metal/electrolyte interfaces in energy applications and beyond.

\section{Background}
The large number of available MLIPs can make it difficult to assess the significance of their design choices. In what follows, we provide an introduction to basic design principles, explain the differences between the models included in our benchmark, and relate them to other models in the literature. Central aspects in our discussion are (1) the body order, related to the model accuracy; (2) the receptive field, i.e., the distance over which atomic interactions are considered by a model; and (3) the computational cost. Finally, we discuss recent advances beyond short-ranged MLIPs to clarify the scope of this work.

\subsection{Locality and descriptors}
The foundation of atomistic simulations is the potential energy surface (PES), i.e., the relationship between atomic positions and chemical species and the corresponding potential energy. MLIPs are designed to be much faster at evaluating the PES than electronic structure methods like DFT, and can hence be used to accelerate molecular dynamics simulations. Unlike classical force fields that use fixed empirical formulae to describe interatomic interactions, MLIPs employ more flexible basis functions. 

To allow for an efficient representation of varying system sizes, most MLIPs decompose the total energy $E$ as a sum of local atomic energies $E_i$ \citep{behler2007generalized}:
\begin{equation} \label{eq:e_decomposition}
    E = \sum_i E_i.
\end{equation}
By this decomposition, the computational cost scales linearly with system size. The local atomic contributions $E_i$ are typically represented as \citep{thiemann2024introduction}
\begin{equation} \label{eq:ei_descriptor}
    E_i = \underline{f}(\mathbf{d}_i)
\end{equation}
where $\mathbf{d}_i=\{d_{i1},d_{i2},...\}$ is a collection of \textit{descriptors} that encode the information in the atomic environment $\mathcal{N}_i$. We denote learnable functions as underlined. The fitting function $\underline{f}$ can be represented by a neural network \citep{behler2007generalized}, a polynomial \citep{drautz2019atomic, shapeev2016moment} or a kernel sum \citep{bartok2010gaussian}. The descriptors representing the environment $\mathcal{N}_i$ are functions of the relative positions $\vec{r}_{ij}=\vec{r}_{j}-\vec{r}_i$ and chemical elements $z_j$ of all neighbors $j$ within a specified cutoff radius $c$. The computational cost of calculating descriptors typically scales with the number of neighbors within the cutoff \citep{drautz2019atomic}. The range within which atoms $j$ can contribute to energy $E_i$ is also known as the \textit{receptive field} of the model. 

The forces on each atom can be predicted either by fitting the force data directly \citep{liao2024equiformerv2improvedequivarianttransformer,neumann2024orbfastscalableneural,gasteiger2024gemnetuniversaldirectionalgraph, bigi2024dark}, or from the gradient \citep{chmiela2017machine, thiemann2024introduction} of the energy with respect to the atomic positions ($\vec{F}_i=-\vec{\nabla}_i E$). Forces obtained as derivatives are conservative, contributing to the stability of MD simulations \citep{chmiela2017machine, bigi2024dark}. All MLIPs considered in our benchmark calculate forces as energy gradients. 

\subsection{Symmetry and body order}
The definition of descriptors $\mathbf{d}_i$ is often motivated by symmetry. The energy is invariant (symmetric) under transformations belonging to the Euclidean group E(3): translations, rotations, and reflections. It is also invariant with respect to permutation, i.e., the order in which we label atoms of the same type. By fitting the energy to descriptors that are E(3)- and permutation-invariant, the model does not need to learn these symmetries from the training data, thereby improving its learning efficiency. 

To satisfy permutation and translation symmetry, the local atomic environment can be represented as a density distribution of neighbor atoms. Descriptors can then be interpreted as coefficients of an expansion of this density in some suitable basis. Concretely, the atomic density centered on atom $i$ is defined as
\begin{equation}
    \rho_{i}(\vec{r}) = \braket{\vec{r}|\rho_i} = \sum_{j\in\mathcal{N}i} g(\vec{r} - \vec{r}_{ij}).
\end{equation}
An index $z$ can be added to $\rho_i$ to define different densities for each chemical species.
This density-based construction provides a unified interpretation of many modern descriptors \citep{musil2021physics}. 
Different models correspond to different choices of $g$ and the basis used to represent $\rho_i$. 
Here we focus on the atomic cluster expansion (ACE), where $g$ is chosen to be a Dirac $\delta$ function \citep{drautz2019atomic}. 

Following the notation of ACE, the density is represented by expanding it in a basis of radial functions $R_{nl}(r)$ and spherical harmonics ${Y}_m^{(l)}(\hat{r})$. The radial basis functions are often learnable (e.g. using a neural network). The expansion coefficients are given by
\begin{equation} \label{eq:ace-a}
    A_{in,m}^{(l)} = \braket{nlm | \rho_{i}}= \sum_{j \in \mathcal{N}_i} \underline{R}_{nl}(r_{ij}) Y_m^{(l)}(\hat{r}_{ij}).
\end{equation}
The expansion is typically truncated at some $l=l_\mathrm{max}$, a hyperparameter of the model.

Ideally, descriptors should also be invariant to rotations and reflections --- together forming the group O(3). Only the spherical harmonics with $l=0$ are O(3)-invariant. However, choosing only $A_{in}^{(0)}$ erases all information except pairwise distances, since $Y^{(0)}_0$ is a constant. Pairwise distances alone cannot provide a complete representation of the atomic density \citep{pozdnyakov2020incompleteness}. 

To retain many-body information, one can instead consider $\nu$-body correlations of the atomic density (see Fig. 5 in Ref.~\onlinecite{musil2021physics}),
\begin{equation}
    \rho^{\otimes\nu}_i = \underbrace{\rho_i \otimes ...\otimes \rho_i}_{\nu \text{ times}}.
\end{equation}
In the basis of the spherical harmonics, these density correlations correspond to tensor products of $A_{in,m}^{(l)}$. Such tensor products also appear in quantum mechanics, where the quantum numbers $l$ and $m$ describe the angular momentum. A tensor product of angular momentum states $\ket{l_1m_1}\otimes\ket{l_2 m_2}=\ket{l_1 m_1;l_2 m_2}$ does not have a unique total angular momentum $L$. Quantities with well-defined $L$ can be constructed by a Clebsch-Gordan contraction (see Section \ref{app:rotation}): 
\begin{equation}
    \ket{l_1l_2LM} = \sum_{m_1m_2} C_{l_1l_2m_1m_2}^{LM} \ket{l_1m_1;l_2m_2},
\end{equation}
where $C_{l_1l_2m_1m_2}^{LM}$ are Clebsch-Gordan coefficients. This operation allows for the construction of tensor product features with combined $L=0$, making them O(3)-invariant. The contraction of higher-order tensor products ($\nu>2$) is described analogously by generalized Clebsch-Gordan coefficients $\mathcal C_\mathbf{lm}^{LM}$ with indices $\mathbf{l}=(l_1,l_2...)$ and $\mathbf{m}=(m_1,m_2,...)$\citep{Yutsis1962}. A set of invariant features can thus be defined as
\begin{equation} \label{eq:b-invariant}
    B_{i\mathbf{n}\mathbf{l},\nu} = \sum_{m_1,..., m_\nu} \mathcal{C}_{\mathbf{l}\mathbf{m}}^{00} A_{in_1,m_1}^{(l_1)}A_{in_2,m_2}^{(l_2)} \cdots A_{in_\nu,m_\nu}^{(l_\nu)}.
\end{equation}

\begin{figure}
    \centering
    \includegraphics[width=\linewidth]{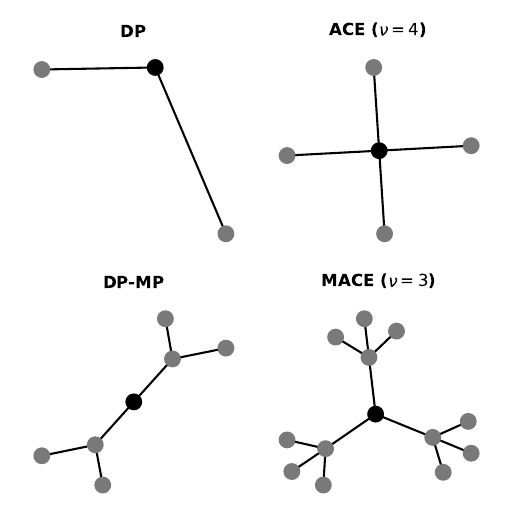}
    \caption{Body order of local models and message-passing models. The black node is the central atom. The graphs represent interactions simultaneously considered in the model. Each message-passing layer makes the graph branch into $\nu$ neighbors, where $\nu$ is the correlation order of the layer. The effective body order is the total number of nodes in the graph. Note that this is a schematic graph illustrating the body order, not a graph representing the full atomistic structure, which connects all atoms within the cutoff radius.}
    \label{fig:graphs}
\end{figure}

The $B$-features provide an expansion of the energy in $\nu$ \citep{drautz2019atomic}. 
The energy $E_i$ can thus be expressed in terms of the descriptors $\mathbf{d}_i=\{B_{i,\mathbf{nl}\nu}\}_{\mathbf{nl}\nu}$ with all $\mathbf{n},\mathbf{l}$ for $\nu$ up to some $\nu_\mathrm{max}$, a hyperparameter. In the implementation of ACE used in this work, $\nu_\mathrm{max}=4$ by default. A closely related concept is the \textit{body order}, which is given by $\nu_\mathrm{max}+1$; it includes the center atom of the environment (Fig.~\ref{fig:graphs}).  A higher body order helps the model to distinguish different configurations, and generally improves accuracy and data efficiency \citep{pozdnyakov2020incompleteness}. 

The expansion coefficients $A_{in,m}^{(l)}$ can be interpreted as components of $(2l+1)$-dimensional vectors $\vec{A}_{in}^{(l)}$, i.e., spherical tensors (Sec. \ref{app:rotation}). Many other MLIPs use cartesian tensors instead. In particular, the DP models use cartesian tensors of rank 0 to 2 \citep{gao2024enhanced}:
\begin{align} 
\begin{split} \label{eq:dp-t-b}
    {T}^{[0]}_{i,n} &= \sum_{j\in\mathcal{N}_i} \underline{R}_n(r_{ij}) \\
    \vec{{T}}^{[1]}_{i,n} &= \sum_{j\in\mathcal{N}_i}  \hat{{r}}_{ij} \underline{R}_n(r_{ij})  \\
    \vec{T}^{[2]}_{i,n} &= \sum_{j\in\mathcal{N}_i} \hat{r}_{ij}\otimes\hat{r}_{ij} \underline{R}_n(r_{ij}).
\end{split}
\end{align}
These tensors are related to the $l=0,1,2$ spherical tensors by a basis transformation. In DP, the radial functions are defined as $\underline{R}_n(r)=s(r)\underline{G}_n(s(r))$, with a smooth cutoff function $s(r)$ and an embedding neural network $\mathbf{\underline{G}}(x)$ that takes a scalar input $x$ and returns $N$ outputs $\underline{G}_1(x),...,\underline{G}_N(x)$. For interactions between different chemical species, different embedding networks $\mathbf{\underline{G}}^{z_iz_j}$ are defined.

The standard DP descriptors are defined as \citep{zhang2018end, gao2024enhanced}
\begin{equation}
    {d}_{i,(n_1,n_2)} = T^{[0]}_{i,n_1}T^{[0]}_{i,n_2} + \vec{T}^{[1]}_{i,n_1} \cdot \vec{T}^{[1]}_{i,n_2},
\end{equation}
where the rank 1 cartesian tensors (vectors) are contracted into scalars by a dot product, analogous to Eq. \ref{eq:b-invariant}. Because these descriptors include only products of order 2 (effectively $\vec{r}_{ij} \cdot \vec{r}_{ik}$, i.e., angular features), the body order is three. To compute the energy $E_i$, the descriptors are stacked into an array $\mathbf{d}_i$, and then fed into a fitting neural network (Eq.~\ref{eq:ei_descriptor}). The nonlinearity of the fitting net can effectively increase the body order, but in an incomplete way; this issue is discussed further in Ref.~\onlinecite{nigam2022unified}. 

In our benchmark, ACE represents high body-order descriptors, similar to the moment tensor potential \citep{shapeev2016moment} and Gaussian moment descriptors \citep{zaverkin2020gaussian}, which implement higher-order correlations with a cartesian basis. DP represents 3-body models, comparable to the atom-centered symmetry functions \citep{behler2007generalized} and the \texttt{aenet} descriptors \citep{artrith2017efficient}.

\subsection{Updating local information by message-passing}
The descriptor-fitting approach described above limits the receptive field to the cutoff radius $c$. We refer to such models as \emph{local}. Message-passing graph neural networks (GNNs) extend the receptive field and further increase the body order by updating information between neighboring atoms.

In a GNN \citep{duval2023hitchhiker}, the atomistic structure is represented as a graph. Atoms are nodes, and edges connect atoms that are within a distance $c$ from one another. Each atom $i$ is assigned an array of updatable node features $\mathbf{h}_i$. In each message-passing layer, atom $i$ receives a \textit{message} from each neighbor $j \in \mathcal{N}_i$, and its features are updated using these messages. Stacking multiple layers allows information to propagate over a distance of $\text{number of layers} \times c$. We therefore refer to these models as \textit{semilocal}. In practice, however, the influence of distant atoms decays with the number of layers \citep{alon2020bottleneck}.

Here we consider a two-layer message-passing neural network based on the DP descriptors, DP-MP \citep{gao2024enhanced}. In the first layer, the node features are initialized with the standard DP descriptors: $\mathbf{h}_i \leftarrow \mathbf{d}_i$. In the second layer, messages are constructed for each neighbor $j$ as
\begin{equation} \label{eq:dp-message}
    \mathbf{m}_{ij} = \text{concat}(\mathbf{\underline{G}}(s(r_{ij})), \mathbf{h}_i, \mathbf{h}_j, \vec{\mathbf{T}}^{[1]}_i \cdot \vec{r}_{ij}, \vec{\mathbf{T}}^{[1]}_j \cdot \vec{r}_{ij} )
\end{equation}
where $\vec{\mathbf{T}}^{[1]}_i$ is the collection of vectors $\vec{T}^{[1]}_{i,n}$ ($n=0,1,...$). The message thus contains information about the distance between $i$ and $j$ and their respective environments, as encoded by the descriptors. The dot products $\vec{\mathbf{T}}^{[1]}_{i} \cdot \vec{r}_{ij}$ express directional information. 

The messages are then used to create updated features $\vec{T}_{i,n}^{[1]}$ and $\vec{T}_{i,n}^{[2]}$ using a message embedding network $\mathbf{\underline{G}}'(\mathbf{m}_{ij})$ that replaces $\mathbf{\underline{G}}(s(r_{ij}))$ in the radial functions of Eq. \ref{eq:dp-t-b}. The six independent components of $\vec{T}_{i,n}^{[2]}$, containing $l=0$ and $l=2$ information, are stacked into a vector. To update the invariant node features, the new features are again contracted by dot products:
\begin{equation} \label{eq:dp-layer2-features}
    {h}_{i,(n_1,n_2)} \leftarrow  \vec{{T}}^{[1]}_{i,n_1} \cdot \vec{{T}}^{[1]}_{i,n_2} + \vec{T}^{[2]}_{i,n_1} \cdot \vec{T}^{[2]}_{i,n_2}.
\end{equation} 
The atomic energies are finally obtained from the last-layer features as $E_i = \underline{f}(\mathbf{h}_i)$, analogous to Eq.~\ref{eq:ei_descriptor}. 

The effective body order of the message-passing model is determined by the highest-order product of single-atom features. Consider a central atom $i$ with neighbors $j$ and $k$. After the first layer, the features $\mathbf{h}_j$ and $\mathbf{h}_k$ are descriptors with $\nu=2$, i.e., their elements are sums of products that each encode information about 2 of their neighbors, which may be different from $i$. By message-passing in the second layer, atom $i$ can therefore accumulate information on up to six surrounding atoms simultaneously. Including the center atom, this corresponds to a body order of 7 (Fig. \ref{fig:graphs}). Other examples of GNNs with $\nu=2$ per layer are M3GNET \citep{chen2022universal} and DimeNet \citep{gasteiger2020directional}. GNNs with $\nu=1$ per layer such as SchNet \citep{schutt2017schnet} thus need more layers to achieve the same body order. 

\subsection{Equivariance: passing directional information} \label{sec:equivariance}
The definition of the message in Eq.~\ref{eq:dp-message} may appear rather ad-hoc, in particular the elements $\vec{\mathbf{T}}^{[1]}_{i,j} \cdot \vec{r}_{ij}$, which convey some kind of directional information. Equivariant GNNs provide a more systematic framework to propagate directional information, by including vectors and higher-rank tensors in the message and node features directly \citep{duval2023hitchhiker}. Equivariance means that rotations of the input lead to corresponding rotations of all intermediate tensors. The symmetry of the energy is still obeyed, as long as the energy is fitted to invariant features (which may be updated by contracting intermediate equivariant features). 

To simplify the representation of tensorial features, many GNNs use spherical tensors. Node features can then be denoted $\vec{\mathbf{H}}^{(0:L_\mathrm{max})}_{i}$ and messages as $\vec{\mathbf{M}}_{i}^{(0:L_\mathrm{max})}$. In this notation \citep{duval2023hitchhiker}, the superscript indicates that tensors with degree $l$ from $0$ to $L_\mathrm{max}$ are included; the vector arrow on top implies that each spherical tensor has $2l+1$ components; and the bold font indicates multiple copies for each degree $l$, called \textit{channels}, typically indexed by $k$. A parity index can also be added to indicate behavior under reflection \citep{batzner20223}.  

In most equivariant GNNs, messages are constructed through a graph convolution, i.e., a weighted aggregation of neighboring node features. Since the features are spherical tensors, this convolution is implemented using weighted Clebsch-Gordan tensor products between a spherical harmonic \emph{filter} and the node features \citep{thomas2018tensor, duval2023hitchhiker}. The general form of this operation is
\begin{equation} \label{eq:cg-tensor-product}
\vec{\mathbf{A}}_{i}^{(0:l_\mathrm{max})}
= \sum_{j\in\mathcal{N}_i} \vec{Y}^{(0:l_\mathrm{max})}(\hat{r}_{ij}) \hspace{0.3em}\underline{\otimes}_{(r_{ij})} \vec{\mathbf{H}}^{(0:L_\mathrm{max})}_{j}
\end{equation}
which considers all possibilities to contract inputs with $l_1\leq l_\mathrm{max}$ and $l_2\leq L_\mathrm{max}$ into outputs $l_3\leq l_\mathrm{max}$. The notation $\underline{\otimes}_{(r_{ij})}$ indicates that the weights of the tensor product are given by a learnable radial basis $\underline{R}_{kl_1l_2l_3}(r_{ij})$. Such weighted tensor products are implemented in packages like \texttt{e3nn} \citep{geiger2022e3nn} and \texttt{cuEquivariance} \citep{nvidia_cuequivariance_2025}. As an example, consider that $\vec{Y}^{(1)}(\hat{r}_{ij})$ is equivalent to the unit vector $\hat{r}_{ij}$, and that the $l=1$ node features may contain descriptor features similar to $\vec{T}_{i,n}^{[1]}$ in Eq.~\ref{eq:dp-t-b}. The tensor product then contains operations similar to $\hat{r}_{ij}\cdot\vec{\mathbf{T}}_{i}^{[1]}$ with scalar outputs $\mathbf{A}_i^{(0)}$, as in the DP message (Eq.~\ref{eq:dp-message}). However, operations leading to higher-degree outputs are now included as well (cross products, outer products, etc.).

In GNNs such as NequIP \citep{batzner20223} and eSEN \citep{fu2025learning}, the output $\vec{\mathbf{A}}_i$ of Eq.~\ref{eq:cg-tensor-product} is used directly as the message, corresponding to pairwise interactions ($\nu=1$) and increasing the body order by one every message-passing layer \citep{batatia2025design}. In MACE \citep{batatia2022mace}, the convolution outputs are only an intermediate, and higher body-order messages are constructed from products of $A$-features. These products are contracted into a well-defined rotation degree $L$ using generalized Clebsch-Gordan coefficients \citep{batatia2022mace}: 
\begin{equation}
    B^{(L)}_{i,k\mathbf{l}\nu,M} = \sum_{\mathbf{m}} \mathcal{C}_{\mathbf{lm}}^{LM} A_{ik,m_1}^{(l_1)}A_{ik,m_2}^{(l_2)} \cdots A_{ik,m_\nu}^{(l_\nu)}.
\end{equation}
This operation is similar to Eq.~\ref{eq:b-invariant}, but now any degree $L$ is allowed for the $B$-features. The messages are then linear combinations of $B$-features, and the node features are updated by a linear combination of the message and the previous node features. In the final layer, the energy is calculated from the $l=0$ node features: $E_i=\underline{f}(\{\mathbf{H}_i^{(0)}\}_\text{(all layers)})$. Because each layer constructs high body-order terms, MACE typically uses only two message-passing layers \citep{batatia2022mace}; with $\nu=3$ per layer, this yields body order 13 (Fig.~\ref{fig:graphs}). The 2-layer GRACE model \citep{bochkarev2024graph} implements a similar idea of passing high-body-order messages based on ACE.

Although equivariant graph neural networks are considered state of the art in terms of accuracy, Clebsch-Gordan tensor products (Eq.~\ref{eq:cg-tensor-product}) are computationally expensive, especially at high $l$ \citep{xie2025price}. The Clebsch-Gordan coefficients form a large sparse tensor with irregular structure, leading to many small matrix multiplications, whereas GPUs are optimized for multiplying large dense matrices. Recent developments focus on exploiting the structure of Clebsch-Gordan coefficients \citep{bharadwaj2025efficient, tan2025high}, using alternative formulations \citep{passaro2023reducing, xie2025price, liao2024equiformerv2improvedequivarianttransformer}, or lifting symmetry constraints altogether \citep{langer2024probing}. 

\subsection{Learning charges and potentials}
Because they lack charge information and are therefore not forced to satisfy electroneutrality in the bulk, short-range MLIPs have been reported to fail in large electrolyte regions \citep{zhang2024molecular, kim2024learning}. To address such problems, there has been significant interest in developing long-range MLIPs.

A simple way to incorporate electrostatics is to learn atomic charges or Wannier centers from local descriptors, and calculate the electrostatic energy in an Ewald sum \citep{zhang2022deep, cheng2025latent, unke2019physnet, gao2022self}. While this usually works for molecular systems, it is not guaranteed to work for a metal electrode, where the charge might depend on electrolyte ions beyond the cutoff radius. Refs.~\onlinecite{zhu2025machine} and \onlinecite{wang2025ion} therefore only apply this method to learn the charges in the electrolyte, and treat the electrode classically. Another approach is to learn the electronegativity from local descriptors and apply charge equilibration to obtain the atomic charges \citep{ko2021fourth, rinaldi2025charge}. However, these charge equilibration methods fail to describe molecules in an electric field \citep{vondrak2025pushing} and metal surfaces \citep{sundararaman2022improving}, making them unsuitable for charged interfaces.

To overcome the limitations of local descriptors, several works have developed long-range descriptors, for example by exploiting reciprocal-space representations and/or transformer architectures \citep{rumiantsev2025learning, ramasubramanian2025reciprocal, unke2021spookynet, grisafi2019incorporating, loche2025fast}. For example, Ref.~\onlinecite{grisafi2024accelerating} used long-range descriptors to learn the charge on a metal electrode, while treating the electrolyte with classical force fields. These approaches may also be able to infer the metal surface charge based on the number of ions in the simulation cell (when trained on constant-charge DFT calculations). 

An alternative strategy is to control the potential or charge explicitly. Some recently developed MLIPs can be used to perform constant-potential molecular dynamics by learning to predict the Fermi level from atomic configurations and the excess electron number obtained from constant-potential DFT calculations \citep{wang2025constant, chen2025grand, chen2025constant}, or by predicting the excess charge given the target potential \citep{chen2023atomistic}. Ref.~\onlinecite{bergmann2025machine} instead proposes learning the work function from implicit-solvation DFT calculations to perform molecular dynamics at constant surface charge.

The aforementioned approaches introduce substantial methodological and computational complexity beyond short-range energy/force MLIPs, and their implementations remain either immature or are not yet broadly available as open-source tools. In this work, we therefore focus on benchmarking short-range MLIPs as the first layer of this complexity, with the goal of clearly identifying the regimes in which purely local models succeed and where explicit long-range or constant-potential treatments become necessary. A systematic assessment of long-range and constant-potential MLIPs is left for future studies. 

\section{Methods}

\subsection{Models}
As summarized in Table~\ref{tab:mlip_overview}, we benchmark five MLIPs: the Deep Potential (DP) and its message-passing variant (DP-MP), the GRACE-1L implementation of ACE, MACE, and eSEN-OC25. DP and DP-MP were run using a JAX implementation. GRACE-1L was used with its TensorFlow interface to LAMMPS. MACE was run with the PyTorch implementation using the ML-IAP LAMMPS interface, with \texttt{cuEquivariance} \citep{nvidia_cuequivariance_2025} acceleration enabled for both training and inference. For eSEN-OC25, we used the pretrained \texttt{esen-sm-conserving-all-oc25} model and ran MD using ASE. All simulations were performed on a single NVIDIA A100 GPU paired with an 18-core Intel Xeon CPU.

\def\arraystretch{1.25}
\begin{table}[tb]
\centering
\caption{Overview of the MLIPs benchmarked in this work, including their implementation, back-end, and molecular dynamics (MD) engine.}
\label{tab:mlip_overview}
\begin{tabular}{lll}
\hline
\textbf{Model} & \textbf{Implementation}\; & \textbf{MD engine} \\
\hline
DP \citep{zhang2018end} & \texttt{deepmd-jax} \citep{deepmdjax2025} & JAX-MD \citep{schoenholz2021jax} \\
DP-MP \citep{gao2024enhanced} & \texttt{deepmd-jax} & JAX-MD \\
GRACE-1L \citep{bochkarev2024graph} & \texttt{gracemaker}\citep{gracetensorpotential2025} & LAMMPS \citep{thompson2022lammps} \\
MACE \citep{batatia2022mace, batatia2025design} & \texttt{mace-torch}\citep{mace2025} & LAMMPS (ML-IAP\citep{SmithEtAl2025}) \\
eSEN-OC25 \citep{fu2025learning, sahoo2025open} & \texttt{fairchem-core}\citep{fairchemcore2025} & ASE \citep{larsen2017atomic} \\
\hline
\end{tabular}
\end{table}

\begin{table*}[t]
\centering
\caption{Model hyperparameters and the resulting receptive field and body order. Hyperparameters are cutoff radius $c$, number of message-passing layers, correlation order per layer $\nu_\mathrm{layer}$, spherical harmonics degree $l_\mathrm{max}$ in the density expansion or convolution filter, and the spherical tensor degree of messages and node features $L_\mathrm{max}$. The value of $l_\mathrm{max}$ for DP-MP is included in brackets because it is only used in the last layer.}
\label{tab:hypers}
\begin{tabular}{lccccc|cc}
\hline
\textbf{Model} 
& {$c$ (\SI{}{\angstrom})} 
& {\#layers} 
& {$\nu_\mathrm{layer}$} 
& {$l_\mathrm{max}$} 
& {$L_\mathrm{max}$}
& {receptive field (\SI{}{\angstrom})}
& {body order}
\\
\hline

DP
& 6
& 1 
& 2 
& 1
& --
& 6
& 3 \\

DP-MP
& 5
& 2 
& 2 
& (2)
& 0
& 10
& 7 \\

GRACE-1L
& 6
& 1
& 4
& 4
& --
& 6
& 5 \\

MACE
& 5
& 2 
& 3
& 3
& 2
& 10
& 13 \\


eSEN-OC25
& 6
& 4 
& 1 
& 2
& 2 
& $\leq24$
& 5 \\

\hline
\end{tabular}
\end{table*}

The most important hyperparameters and the resulting receptive field and effective body order are summarized in Table~\ref{tab:hypers}. 
For MACE, node features and messages consist of 64 scalar features, 64 vector features, and 64 even-parity tensor features with degree $l=2$. 
Although the eSEN model has four message-passing layers, some long-range information may be lost \citep{alon2020bottleneck}, as indicated in Table~\ref{tab:hypers} by the $\leq$ symbol. 
Additional details on hyperparameters and training procedures for all models are provided in Section \ref{si:training_details}.

\subsection{Data}
The data consists of Au/water interfaces with 0, 1, 2, 3, or 4 explicitly solvated sodium ions (\ce{Na^+}). The supercell contains a $6 \times 4$ Au(111) slab with a thickness of four layers and lattice constant $a = \SI{4.20}{\angstrom}$, several water layers with a total thickness of approximately \SI{20}{\angstrom}, and a \SI{20}{\angstrom} vacuum region. The structures contain between 376 and 384 atoms.

With these systems, three datasets were constructed:
\begin{itemize}
    \item a \emph{mixed dataset} of 3500 structures, with equal representation of systems containing 0 to 4 sodium ions (700 structures per ion count);
    \item an \emph{neutral-surface dataset} containing 1200 Au/water structures without solvated ions;
    \item a \emph{negative-surface dataset} containing 2600 structures with three solvated \ce{Na^+} ions, thus representing an effective negative surface charge of $\SI{-0.016}{e/\angstrom^2}\approx\SI{-26}{\micro\coulomb/\centi\meter^2}$.
\end{itemize}
We refer to the latter two datasets as `specific' datasets because they only contain systems with a specific surface charge. 

These datasets were constructed by taking an initial dataset from Ref.~\onlinecite{ankit2025inpreparation} and generating additional structures by active learning. The active learning procedure, which uses the minimum set cover sampling algorithm of Refs.~\onlinecite{schwalbekoda2025information} and \onlinecite{yu2025maximizing}, is described in detail in the SI, S-II. In addition, two independent test sets were taken from Ref.~\onlinecite{ankit2025inpreparation}: one ion-free test set with 2690 structures, and one three-ion test set with 2463 structures. 

\subsection{DFT calculations}
DFT calculations to label the training data were performed with the Vienna Ab Initio Simulation Package (VASP) version 5.4.4 \citep{kresse1993ab, kresse1994ab, kresse1996efficiency, kresse1996efficient, kresse1999ultrasoft}. Consistent with Ref.~\onlinecite{ankit2025inpreparation}, the RPBE functional \citep{hammer1999improved} was used in combination with the D3 dispersion correction with zero damping \citep{grimme2010consistent}. The plane-wave cutoff energy was set to \SI{450}{\electronvolt}, and the Brillouin zone was sampled with a $2 \times 3 \times 1$ Monkhorst-Pack \citep{monkhorst1976special} $k$-point grid. Interactions of the slab dipole with its periodic images were avoided by a dipole correction; the dipole was calculated with respect to the mean z-coordinate of all atoms. The convergence criterion of the self-consistent cycle was \SI{1e-6}{\electronvolt} for the change in energy. Gaussian smearing with $\sigma=0.1$ was used, and the energy extrapolated to $\sigma\to0$ was used as energy label for the training data. The OC25 dataset uses very similar DFT parameters \citep{sahoo2025open}. 
DFT-MD reference trajectories for the neutral Au/water system were obtained from Ref.~\onlinecite{ankit2025inpreparation}.

\subsection{Molecular dynamics simulations}
All JAX-MD and LAMMPS simulations were performed in the NVT ensemble at \SI{300}{\kelvin}. The starting configurations were obtained from trajectories driven by MACE that were stable and equilibrated for at least \SI{100}{\pico\second}. Momenta were initialized with a Maxwell-Boltzmann distribution. A Nosé-Hoover chain \citep{nose1984unified, hoover1985canonical, martyna1992nose, martyna1996explicit, tuckerman2006liouville} thermostat with chain length 3 was used. JAX-MD simulations used a damping parameter of \SI{100}{\femto\second} (default), LAMMPS simulations a damping parameter of \SI{50}{\femto\second}. The thermostat was applied to all atoms except the fixed atoms in the Au slab. The timestep was set to \SI{0.5}{\femto\second}, and frames were saved every \SI{10}{\femto\second}. A comparison of the JAX-MD and LAMMPS thermostats can be found in Fig.~S5.

The bottom two layers of the Au slab were fixed. Because fixed-atom constraints are not natively available in \texttt{deepmd-jax}, this functionality was implemented in a fork \citep{dekam_deepmdjax2025}.

The eSEN-OC25 simulations used the same settings as described above, with a few differences: a Langevin thermostat was used with friction coefficient \SI{0.0002}{\femto\second^{-1}}, and the timestep was \SI{1.0}{\femto\second}.

\begin{figure*}
    \centering
    \includegraphics[width=\linewidth]{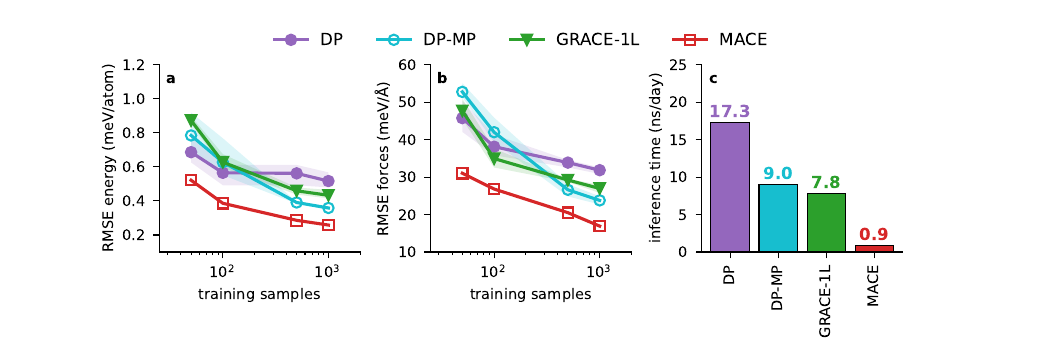}
    \caption{Accuracy and performance of the deep potential (DP), deep potential with message-passing (DP-MP), 1-layer GRACE (GRACE-1L) and MACE using the neutral-surface dataset. (a-b): root-mean-squared error (RMSE) of the energy (a) and force components (b) for models trained on a varying number of structures and evaluated on a test set containing 200 structures. Shaded areas indicate the minimum and maximum results over three different data splits; the markers indicate the mean. MACE was repeated only once. (c) Molecular dynamics speed for an Au/water interface with 384 atoms and a \SI{0.5}{\femto\second} timestep on an NVIDIA A100 GPU.  }
    \label{fig:rmse-perf}
\end{figure*}

\subsection{Trajectory analysis}
Trajectory analysis was performed with the package MDAnalysis \citep{gowers2019mdanalysis, michaud2011mdanalysis} and custom analysis code.\citep{dekam_watanalysis2025} Density profiles were computed from atom position histograms along the trajectory, using the oxygen coordinates for water. Dipole orientation profiles were obtained by computing normalized water bisector vectors $\vec{p}^i$ and weighting water position histograms by their z-components, $p_z^i=\cos\theta_i$, where $\theta_i$ is the angle between the bisector and the surface normal. The total dipole in the z-direction was defined as $P_z=\sum_i \cos\theta_i$, and histograms of $P_z$ were computed over the trajectory. Note that the water dipoles are defined purely geometrically here, so $P_z$ is a dimensionless quantity. $P_z$ is correlated with but not identical to the slab dipole obtained from DFT calculations (Fig. \ref{fig:dipoles}).

\subsection{Uncertainty estimation}
Two sources of uncertainty were considered for the spatial profiles: finite trajectory length, and uncertainty associated with MLIP training.

Statistical uncertainty from finite trajectory length was estimated by block averaging. Each trajectory was divided into $N_\mathrm{block}=10$ blocks, and density and dipole orientation profiles were computed for each block. The standard error was estimated as $\sigma_\mathrm{block}/\sqrt{N_\mathrm{block}}$, yielding a 95\% confidence interval of approximately $2\sigma_\mathrm{block}/\sqrt{N_\mathrm{block}}$. For water density and dipole orientation profiles from \SI{1}{\nano\second} trajectories or more, this interval is smaller than the plotted line thickness; for ion density profiles it will be shown explicitly.

The uncertainty associated with MLIP training was estimated for ion density profiles, where substantial differences were observed between MLIP architectures. Following the committee model approach outlined in Refs.~\onlinecite{musil2019fast, imbalzano2021uncertainty}, $M=5$ models were trained on different subsamples of the same dataset. From the mixed dataset (3500 structures), a validation set of 500 structures was taken first, followed by sampling of five training sets of size 2500, always containing equal numbers of systems with 0–4 ions. From the three-ion dataset (2600 structures), 250 structures were held out for validation and five training sets of size 2250 were sampled.
Because the datasets used for training different models within a committee mostly contain the same structures, their outputs are not statistically independent. To correct for the resulting underestimation of uncertainty, Refs.~\onlinecite{musil2019fast, imbalzano2021uncertainty} propose a scaling factor $\alpha$ computed from validation set errors. Further details and the resulting values of $\alpha$ are reported in the SI in Section \ref{si:reweighting}.

Ion density profiles were obtained from a trajectory generated with a single committee member. Profiles for the remaining models were then estimated by histogram reweighting using the cumulant expansion approximation proposed in Ref.~\onlinecite{imbalzano2021uncertainty} (see \ref{si:reweighting}). The uncertainty associated with MLIP training was taken as the standard deviation over $M=5$ reweighted profiles, scaled by $\alpha$.

\section{Results and discussion}
In the literature, the accuracy of MLIPs is often judged by their energy and force predictions on a test dataset. We start our discussion with such a comparison for the DP, DP-MP, GRACE and MACE models, and discuss the tradeoff between model accuracy and computational cost. We then turn to more practical benchmarks, and investigate whether the models are able to drive stable and accurate MD simulations. We discuss the impact of including different surface charge states in the training set on the obtained water structure and the behavior of solvated ions. Finally, we use these observations to rationalize the behavior of the eSEN-OC25 model.

\subsection{Model accuracy and computational cost}
The accuracy of energy and force predictions is studied as a function of training set size. The training sets of different sizes are sampled from the neutral-surface Au/water dataset. Figure \ref{fig:rmse-perf}a and b show the root-mean-squared error (RMSE) on a test set for energy and force components, respectively. The MACE models consistently show the lowest errors, with MACE trained on 50 structures already outperforming the DP model trained on 1000 structures. Regarding prediction accuracy, the MACE models are therefore much more data-efficient than the DP models. The message-passing DP model shows high errors when trained on 50 to 100 structures, but surpasses the accuracy of the local models when trained on 500 to 1000 structures. 

For models trained on the largest datasets, accuracy increases with receptive field and body order (Table \ref{tab:hypers}). The message-passing models achieve an RMSE that is about 1.5-2 times smaller than that of the local models. As for the local models, GRACE-1L is comparable in accuracy to DP for small training sets, but is more accurate for larger training sets, likely due to its higher body order. Ref.~\onlinecite{pozdnyakov2020incompleteness} previously highlighted the importance of a high body order; their experiments with a \ce{CH4} dataset show improved accuracy with increasing body order as well.

The computational cost is another crucial aspect that has to be taken into account when choosing an MLIP architecture. The time needed to train the models on the 1000-sample training set with an A100 GPU varies from an hour for DP to four hours for equivariant MACE. However, training time is strongly dependent on details of the optimization procedure, such as the learning rate scheduler. Moreover, training is usually one-off, in which case the computational cost is negligible compared to that of inference. We therefore focus on inference efficiency in the following. 

The inference time of MD simulations (ns/day) for the various models is shown in Figure \ref{fig:rmse-perf}c. The general trend is that more accurate models are slower. The DP model is almost 20 times faster than MACE, and DP-MP and GRACE-1L are both about 8-9 times faster. We can identify several architectural and implementation choices that cause these differences in evaluation time. Firstly, DP and DP-MP make use of an algorithm that compresses dense neural networks into polynomials after training \citep{lu2022dp}, greatly improving the evaluation speed. Other examples of implementations using polynomials to achieve high performance are \texttt{PACE} \citep{lysogorskiy2021performant} and the moment tensor potential \citep{novikov2020mlip}. In addition, equivariant graph neural networks like MACE generally rely on computationally expensive tensor product operations. However, the poor scaling in $l$ of these operations should not be a problem when using scalar messages. In Fig.~\ref{fig:rmse-cost-si}, we show that even with scalar messages, MACE is still a factor of 4 slower than GRACE-1L, whereas only a factor of 2 would be expected for two message-passing layers, suggesting that overhead is the main bottleneck. We also show that 2-layer equivariant GRACE (GRACE-2L) is much faster, reaching \SI{6.6}{\nano\second/day}, while maintaining an accuracy comparable to MACE. As discussed in Sec.~\ref{sec:equivariance}, MACE and GRACE-2L have a similar architecture. The substantial difference in evaluation speed emphasizes that practical implementation can strongly affect the computational cost of MLIP-driven simulations.

\subsection{Stability and accuracy with limited training data}
As discussed in Ref.~\onlinecite{fu2022forces}, accurate energies and forces alone do not guarantee stable molecular dynamics over long timescales. To assess model stability in the low-data regime, we ran \SI{100}{\pico\second} NVT simulations with models trained on $n=50$ structures. Fig.~S8 shows the potential energy along these trajectories. For DP and DP-MP, two out of three trajectories depart significantly from the initial energy, indicating instability. GRACE-1L and MACE are much more stable (even for \SI{1}{\nano\second}, see Fig.~\ref{fig:trajectory-energy}).

In Fig.~\ref{fig:pes-cut}, we show that the learned potential energy surface (PES) is much rougher for DP and DP-MP than for GRACE-1L and MACE when trained on 50 structures. The difference in PES roughness may in part be related to different radial basis functions \citep{fu2025learning}: DP models use only dense neural networks \citep{gao2024enhanced} (Eq.~\ref{eq:dp-t-b}) whereas GRACE and MACE use Chebyshev polynomials \citep{lysogorskiy2025graph} and Bessel functions \citep{batatia2022mace} fed into a dense neural network. A rough PES leads to large force predictions, which can destabilize the dynamics and cause the system to explore unphysical configurations that are not represented by the training data.

\begin{figure}
    \centering
    \includegraphics[width=\linewidth]{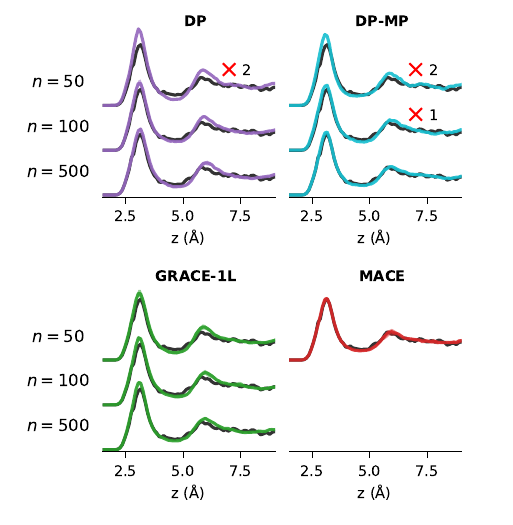}
    \caption{Water density profiles in the direction perpendicular to the gold surface from 100ps trajectories driven by models trained on different amounts of training data, $n$, for the neutral-surface dataset. For each model and training set, the mean of three repeats is shown (in color) with models trained on different subsets of the data and a different initial configuration. The min/max deviation over the three repeats is within the plotted line thickness. If trajectories were unstable, they are not included, and instead the number of unstable trajectories is indicated next to the red $\times$ in the density profile plots. Black lines indicate the mean over three \SI{10}{\pico\second} DFT-MD trajectories.}
    \label{fig:data-efficiency}
\end{figure}

The water density obtained from simulations with training set sizes of 50, 100, and 500 is shown in Fig.~\ref{fig:data-efficiency} (corresponding dipole orientation profiles are shown in Fig.~\ref{fig:water-orientation-si}). Local models deviate slightly from the DFT-MD reference in the second density peak around \SI{6}{\angstrom}, whereas message-passing models agree closely with DFT-MD. This result suggests that a receptive field of \SI{6}{\angstrom} is insufficient to accurately capture all water-surface interactions. Overall, all MLIPs yield reasonable simulations of water near a neutral gold surface, with semilocal models performing best, particularly further from the surface where longer-range interactions become more relevant. 

MACE faithfully reproduces the density profile when trained on 50 structures, again highlighting its data efficiency. The ability to run molecular dynamics reliably with a model trained on only 50 structures makes it feasible to label the training data with more accurate, costly QM calculations. In this way, data-efficient MLIPs like MACE may enable the exploration of metal/water interfaces with higher levels of theory. 

\begin{figure*}
    \centering
    \includegraphics[width=\linewidth]{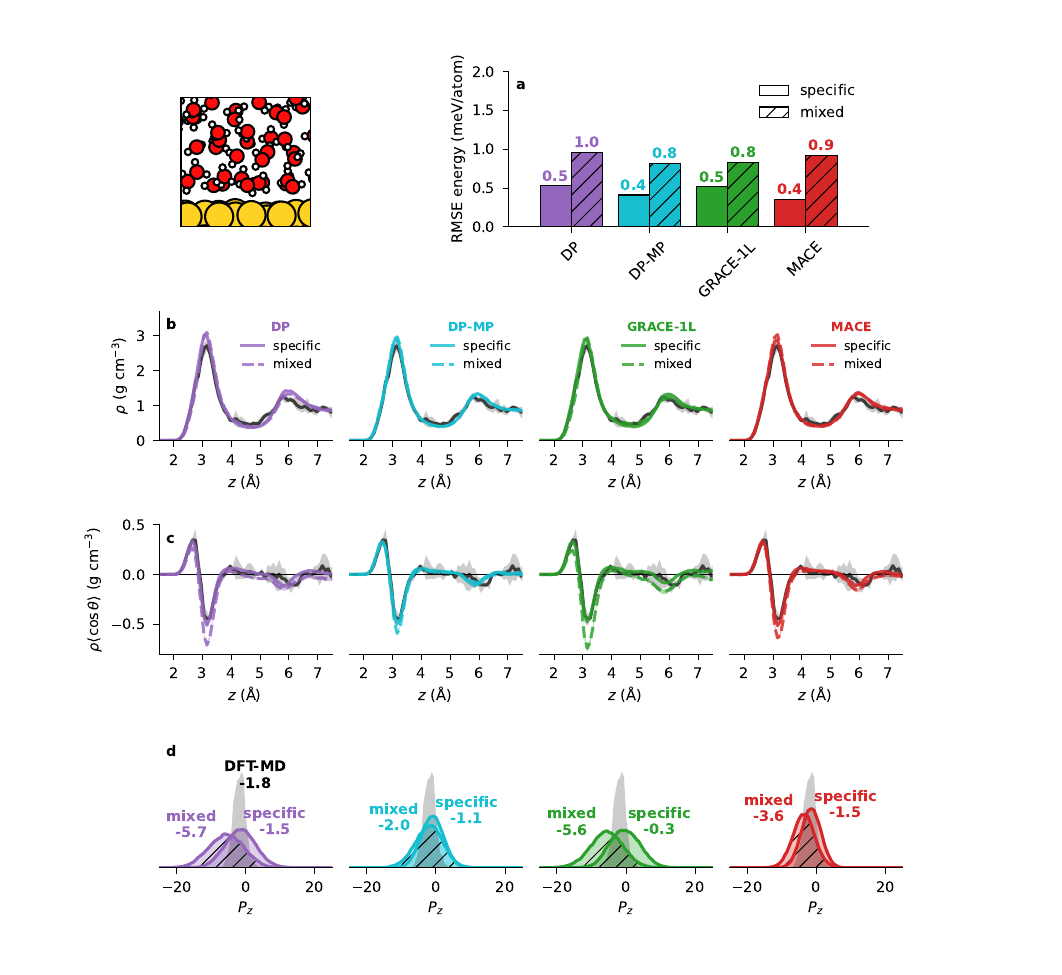}
    \caption{Water structure at a neutral gold surface for models trained only on the target system (`specific') vs. models trained on a dataset with differently charged surfaces (`mixed'). (a) Energy RMSE on the neutral Au/water test set with 2690 structures. (b) Water density profiles and (c) dipole orientation profiles obtained from \SI{1}{\nano\second} molecular dynamics trajectories. $\theta$ is the angle between the water bisector and the surface normal. Black line shows the mean and min/max spread over three \SI{10}{\pico\second} DFT-MD simulations. (d) Total dipole ($P_z$) histograms from molecular dynamics. The mean of the distribution is indicated. The combined result from three \SI{10}{\pico\second} DFT-MD trajectories is shown in gray and reproduced in all plots.}
    \label{fig:0na-water}
\end{figure*}

\begin{figure*}
    \centering
    \includegraphics[width=\linewidth]{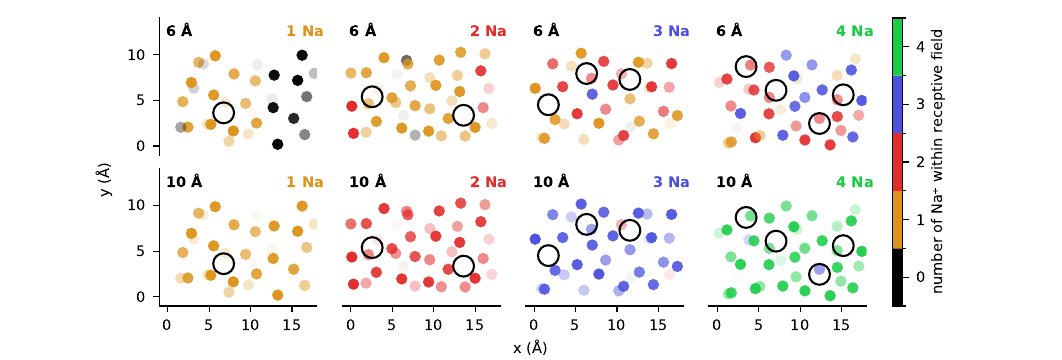}
    \caption{Number of \ce{Na^+} ions within the receptive field of interfacial water molecules. Each subplot shows the $x,y$ positions of oxygen atoms from frames of MACE trajectories; marker transparency depicts the distance $\Delta z$ from the surface with $\Delta z = \SI{3}{\angstrom}$ corresponding to the least transparent markers and $\Delta z = \SI{7.5}{\angstrom}$ corresponding to the most transparent markers. Marker color indicates the number of \ce{Na^+} ions within the receptive field of each oxygen atom. Open markers denote the positions of the \ce{Na^+} ions. The receptive field radius and the total number of sodium ions in the simulation cell are indicated in the top-left and top-right corners of each subplot, respectively.}
    \label{fig:cutoffs}
\end{figure*}

\subsection{Water structure and the effect of mixed training sets}
Having established that MLIP simulations of water at a neutral surface are reasonably reliable, we now consider the water structure in more detail. Specifically, we study the impact of adding differently charged interfaces to the training set by comparing \emph{specific} models, trained only on the target system, with \emph{mixed} models, trained on interfaces with different surface charges in addition to the targeted surface charge. We first analyze the neutral-surface system, followed by a negatively charged surface with three \ce{Na^+} ions.

On the neutral Au/water test set, the energy RMSE of the charge-specific models is lower by roughly a factor of two compared to the mixed-dataset models (Fig.~\ref{fig:0na-water}a), whereas the corresponding difference in force RMSE is comparatively small, only about 10–20\% (Fig.~\ref{fig:ion-water-force-rmse}). Figures \ref{fig:0na-water}b-d show how these errors manifest themselves in the MD results. Although the water density profiles (Fig.~\ref{fig:0na-water}b) from \SI{1}{\nano\second} MD simulations are still similar across the different models and agree well with DFT-MD, differences arise in the water orientation distributions (Fig.~\ref{fig:0na-water}c). The models trained on the mixed dataset exhibit an anomalously large negative peak in the orientation profile, indicating that they favor water to be in the H-down orientation too strongly. This error is larger for the local models (DP and GRACE-1L) than for the semilocal models (DP-MP and MACE). As shown in the previous section, the specific models all match the DFT-MD reference rather well, although for all models, the small negative feature in the orientation profile around \SI{6}{\angstrom} is shifted slightly closer to the surface.

The distributions of the total dipole orientation $P_z$ shown in Fig.~\ref{fig:0na-water}d depict the variation of the water orientation over time. Local models yield much broader distributions than message-passing models. The message-passing models approach the DFT-MD distribution most closely. The difference between the specific and mixed-dataset models is also smaller for the message-passing models (DP-MP and MACE) than for the local models (DP and GRACE-1L). 

For the system with 3 \ce{Na^+} ions and a corresponding negative surface charge (Fig.~\ref{fig:ions}), the mixed-dataset models now favor the H-down orientation \emph{less} than the specific models, indicated by a smaller negative peak in the dipole orientation profile and a more positive $P_z$ distribution. While a DFT-MD reference is not available due to the slow equilibration of solvated ions, the specific models agree well across different architectures. Fig.~\ref{fig:water-dynamics-dip-corr} shows similar results for the interfacial water dynamics, with mixed-dataset models predicting a faster decay of the water dipole autocorrelation function near a charged surface than the specific models.

The incorrect water orientation predicted by mixed-data models indicates underfitting. Since the surface charge is determined by the number of ions in the simulation cell, all interfacial atoms must have access to this global information in order to reproduce the correct behavior. Figure \ref{fig:cutoffs} illustrates how effectively interfacial water molecules can access this information, depending on the model's receptive field. For local models ($c=\SI{6}{\angstrom}$), a large fraction of water molecules does not have the correct number of ions within the receptive field. As a result, the model learns an average behavior across all surface charges in the training set, favoring H-down water orientation too strongly at a neutral surface and too weakly at a negative surface. 
For the message-passing models, the receptive field of \SI{10}{\angstrom} covers the interface almost entirely for the simulation cells considered here. Consequently, these semilocal models should better distinguish the behavior at differently charged surfaces when trained on the mixed dataset. The remaining errors may arise either because the receptive field does not quite cover the entire interfacial water layer, or because the models are not expressive enough to encode the total number of \ce{Na^+} ions. A distant \ce{Na^+} ion causes a subtle change in the local geometry, but can cause significantly different behavior by altering the global surface charge.

\begin{figure*}
    \centering
    \includegraphics[width=\linewidth]{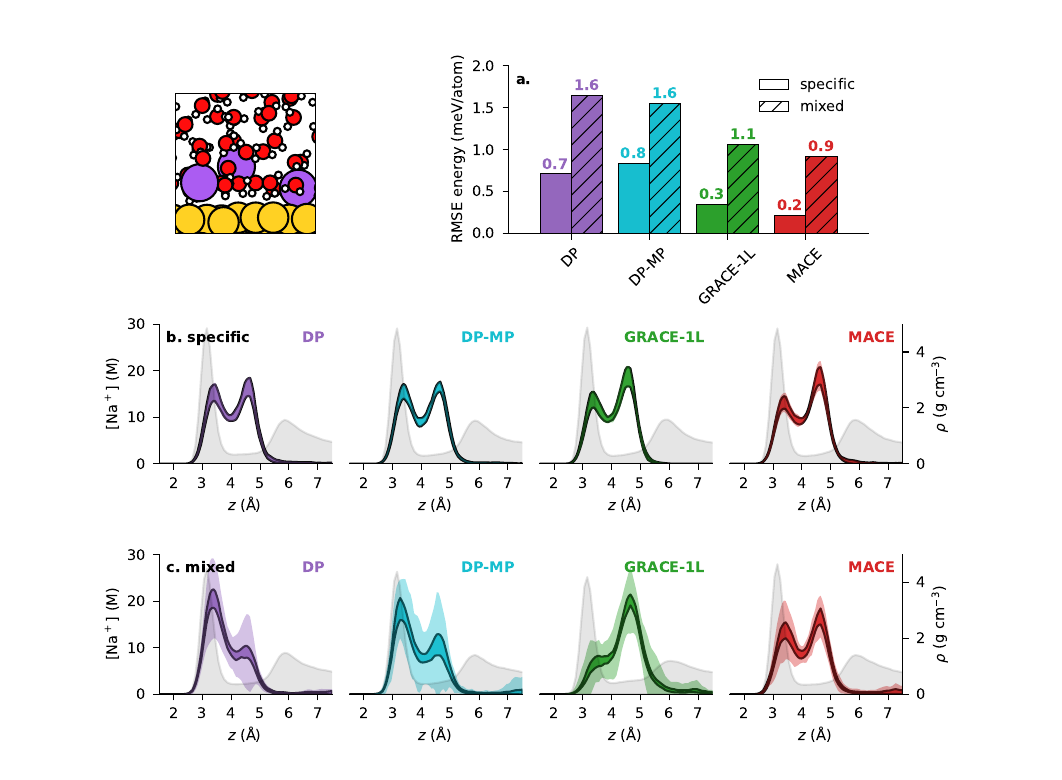}
    \caption{Sodium ion density profiles in the Au/water/3\ce{Na^+} system, comparing models trained only on the target system (`specific') vs. models trained on a dataset with differently charged surfaces (`mixed'). (a) Energy RMSE on the Au/water/3\ce{Na^+} test set with 2463 structures. (b-c) Sodium ion concentration profiles (colored curves) along the surface normal, overlaid with the water density profile (shaded grey), obtained from \SI{2}{\nano\second} molecular dynamics simulations using the specific models (b) and the mixed-dataset models (c). The darker-colored band around each ion profile denotes the \SI{95}{\percent} confidence interval arising from finite sampling. The lighter shaded region indicates the corrected standard deviation $\alpha\sigma$ related to the composition of the training set as estimated by the reweighting procedure.}
    \label{fig:ion-densities}
\end{figure*}

The broadening of the distributions of the total dipole moment $P_z$ is likely a consequence of the absence of long-range electrostatics. In short-range MLIPs, water molecules only `see' other molecules within their receptive field, so there is no energetic penalty for developing extremely positive or negative polarizations. Anomalous water ordering has also been reported for MLIP simulations of water-vacuum interfaces \citep{gao2022self}. Due to their larger receptive field, message-passing models capture electrostatic interactions more effectively than local models, resulting in more narrow $P_z$ distributions. 

In conclusion, errors in the average water orientation arise from the difficulty of distinguishing surface charge states when training on mixed datasets with local descriptors. In contrast, fluctuations in the total water dipole $P_z$ are related to the lack of long-range electrostatics, which can cause anomalous water ordering. As shown in Fig.~\ref{fig:dipoles}, a change in $P_z$ of 5 units corresponds to a change in work function of \SI{\sim 1}{\electronvolt}. Errors in water orientation will thus significantly affect work functions and electrode potentials \citep{le2017determining} calculated from MLIP-driven MD frames, as well as catalytic properties such as reaction barriers \citep{shah2024platinum, tian2025electrochemical}. Other than that, MLIPs trained specifically on the target system do provide a reasonably reliable description of averaged properties of interfacial water, and are consistent across architectures. 

\subsection{Ion density profiles}
\begin{figure}[t]
    \centering
    \includegraphics[width=\linewidth]{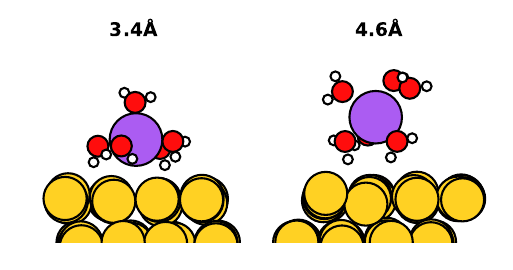}
    \caption{Representative examples of \ce{Na^+} solvation structures. Left: ion within the first water layer. Right: ion just above the first water layer. Ion-surface distances are indicated.}
    \label{fig:solvation}
\end{figure}

The ion density profiles in Fig.~\ref{fig:ion-densities} show that the choice of MLIP architecture and training dataset can lead to different ion distributions. Two favorable positions for solvated \ce{Na^+} ions are observed: one at approximately \SI{3.4}{\angstrom} from the surface, and one around \SI{4.6}{\angstrom}. Representative solvation structures are shown in Fig.~\ref{fig:solvation}. Ions at \SI{3.4}{\angstrom} are partially desolvated and reside within the first water layer, while ions at \SI{4.6}{\angstrom} show a full first solvation shell and reside above the first water layer. GRACE-1L and MACE trained on a specific surface charge slightly favor the position above the first water layer, while DP and DP-MP populate both positions more evenly (Fig.~\ref{fig:ion-densities}b). Although no DFT-MD reference is available for direct validation, the predictions by MACE and GRACE-1L are likely (though not guaranteed) to be the most accurate, given their low energy errors (Fig.~\ref{fig:ion-densities}a). A possible reason is that ion hydration is better described by the higher body-order ACE descriptor. Interestingly, local and semilocal models based on the same descriptor yield nearly identical ion distributions, suggesting that the absence of explicit long-range electrostatics does not affect the resulting ion distribution. Although the peak population is expected to depend on surface charge \citep{moss2025if, alfarano2021stripping}, MLIPs trained on charge-specific data are thus able to learn the effective interactions corresponding to that particular surface charge.

When comparing models trained on mixed datasets, much larger differences arise. Fig.~\ref{fig:ion-densities}c shows that trajectories obtained with different architectures vary significantly: on average, the mixed-data DP and DP-MP models place most ions within the first water layer, whereas mixed-data GRACE-1L, and to a lesser extent MACE, favor the ion position above the first water layer.

To rationalize the differences in ion profiles obtained with different mixed-dataset models, one could consider comparing the test error --- a measure of `noise' in the predicted PES --- to free energy differences. In the canonical ensemble, the free energy surface of the ions is proportional to the logarithm of the equilibrium ion density. A twofold difference in ion concentrations at \SI{3.4}{\angstrom} and \SI{4.6}{\angstrom} corresponds to a free energy difference of $\Delta F\approx k_\mathrm{B}T\log2 \approx \SI{18}{\milli\electronvolt}$. This energy scale is small compared to the test RMSEs in Fig.~\ref{fig:ion-densities}a (e.g., around \SI{70}{\milli\electronvolt} per structure for specific MACE and $3 \times 10^2$ \si{\milli\electronvolt} for mixed-data MACE). Nevertheless, test errors that exceed this free energy difference do not lead to inconsistencies in the ion profiles from the specific models, whereas they do for the mixed-dataset models. This inconsistency highlights that test-set errors alone are not enough to assess the reliability of the results.

To understand the impact of the energy error on the ion density profile more rigorously, we apply the uncertainty estimation procedure described in the Methods section. The lighter shaded regions in Figures \ref{fig:ion-densities}a and b represent the estimated prediction standard deviation, $\alpha\sigma$. The uncertainty is much smaller for the specific models (Fig.~\ref{fig:ion-densities}a) than for the mixed-dataset models (Fig.~\ref{fig:ion-densities}b). The large uncertainty associated with the mixed-dataset models indicates a high sensitivity to changes in the training set composition, in line with the expectation that model predictions depend on the different surface charges included in the training set. This result also demonstrates that uncertainty estimation is a valuable tool to verify the outcome of an MLIP-driven simulation: when a short-range model cannot fully parametrize the global surface charge, the estimated uncertainty is large. Interestingly, a low test RMSE does not necessarily imply low uncertainty. For example, the specific MACE model exhibits slightly higher uncertainty than specific DP-MP, despite having a considerably lower test error.

Including differently charged surfaces in the training set thus leads to unreliable results regarding the equilibrium behavior of solvated ions. These errors result from the difficulty in parameterizing the global surface charge, not from the absence of explicit long-range electrostatics. Message-passing models, in particular MACE, show the smallest deviation when trained on a mixed dataset. However, despite their reasonably large receptive field (see Figure~\ref{fig:cutoffs}), they do not seem to be capable of fully parameterizing the global surface charge. On the other hand, even short-range MLIPs seem to  give reliable results for the ion positions in the Helmholtz layer, as long as they are trained on a single surface charge state.

\begin{figure*}
    \centering
    \includegraphics[width=\linewidth]{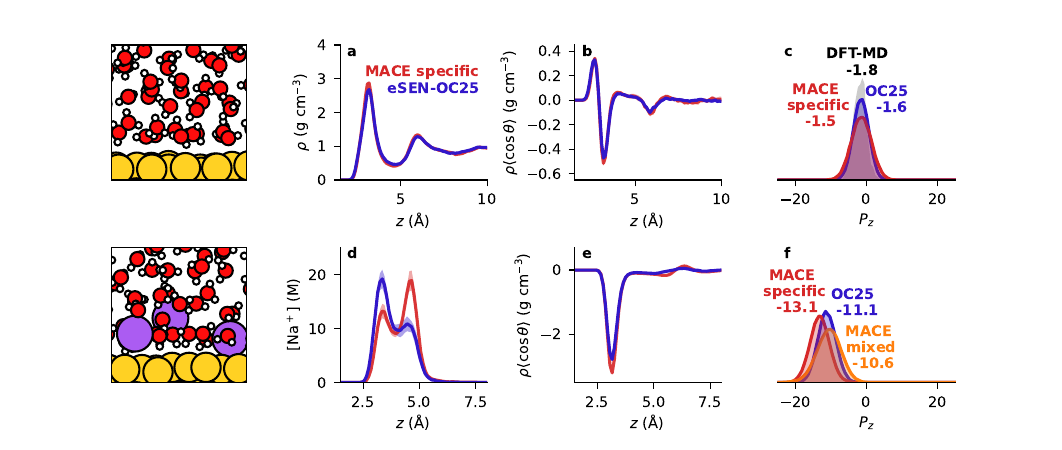}
    \caption{Molecular dynamics with the eSEN-OC25 model on the Au/water system (a-c) and the Au/water/3\ce{Na^+} system (d-f), compared to the MACE results shown previously. eSEN-OC25 trajectories were run for \SI{1}{\nano\second}. The \SI{95}{\percent} confidence interval based on the uncertainty due to finite trajectory length is indicated by the shaded region around the ion density profiles (and falls within the plotted line width for the water density and orientation).}
    \label{fig:oc25}
\end{figure*}

\subsection{The OC25 dataset}
The inability of short-range MLIPs to infer the global surface charge may also affect foundational models trained on large datasets that include a wide range of different surface charges, such as the OC25 dataset from Ref.~\onlinecite{sahoo2025open}. To explore such underfitting effects, Figure~\ref{fig:oc25} compares the eSEN-OC25 model to MACE models trained on specific datasets, which we considered to be the most accurate in the previous section. 

The water density profile, dipole orientation, and total dipole $P_z$ for the neutral-surface Au/water system (Fig.~\ref{fig:oc25}a-c) predicted by eSEN-OC25 agree closely with those predicted by the specific MACE model. Further, we find that the total dipole distribution obtained from the eSEN-OC25 model simulations are slightly narrower and closer to those obtained from DFT-MD simulations,  as compared to the distribution predicted by the specific MACE model. The narrower distribution is likely due to the eSEN model having 4 message-passing layers, resulting in a larger receptive field, which enables the learning of long(er)-range interactions.

For the Au/water/3\ce{Na^+} system (Fig.~\ref{fig:oc25}d-f), however, we observe larger differences between the OC25 model and specific MACE model. The large first peak in the ion density profile indicates that the eSEN-OC25 model favors ions to be within the first water layer, similar to the mixed-data DP models in Fig.~\ref{fig:ion-densities}c. The peak in the dipole orientation profile (Fig.~\ref{fig:oc25}e) is slightly smaller than for the charge-specific MACE model, consistent with Fig.~\ref{fig:ions} in the SI, where the mixed-data models also predict weaker water orientation. This observation is confirmed by the total dipole histograms in Fig.~\ref{fig:oc25}f: the OC25 model predicts a slightly less negative total dipole compared to the specific MACE model. 

We attribute the above-mentioned differences to the inclusion of systems with varying surface charges in the OC25 dataset (ranging from -80 to +\SI{60}{\micro\coulomb/\centi\meter^2}). Because the short-range MLIP struggles to distinguish global surface charge, model training on the dataset with several neutral/weakly charged surfaces likely causes the differences observed in Fig.~\ref{fig:oc25}d-f. Nevertheless, Fig.~\ref{fig:oc25}f shows that the OC25 total dipole distribution is closer to the specific MACE result than that of the mixed-data MACE model. We posit that this improvement is due to the larger receptive field of the OC25 model and larger number of parameters (ca. 6M, compared to 0.2-0.3M for a typical MACE model). 

\section{Conclusion and outlook}
In this work, we benchmarked commonly used short-range MLIPs on various properties of metal/electrolyte interfaces, such as the dipole orientation of interfacial water molecules and the density distribution of solvated ions. When trained on DFT calculations of metal/water interfaces that are charged explicitly by including various numbers of solvated ions, short-range MLIPs struggle to capture global surface charge effects. Because the global surface charge is set by the number of counterions in the simulation cell, all atoms in the interfacial region need to `see' these counterions. Local models with a receptive field of $\sim\SI{6}{\angstrom}$ therefore cannot distinguish between surface charge states when some counterions are beyond the cutoff radius. When trained on mixed-charge datasets, they show significant errors in the predicted water orientation and the distribution of solvated ions. Message-passing MLIPs such as MACE have a larger receptive field of $\sim\SI{10}{\angstrom}$, which improves their performance for mixed-charge datasets. However, even when the receptive field spans almost the entire interfacial region, dipole orientation and ion distributions can be inaccurate, emphasizing the difficulty of encoding global information in a local descriptor. Models trained across different surface charges should, therefore, be used with caution.

On the other hand, our results demonstrate that models trained specifically on the targeted system with a single global surface charge are generally reliable for equilibrium properties of the Helmholtz layer (less than \SI{1}{\nano\meter} from the surface). Local and semilocal (message-passing) models show consistent trajectory averages, suggesting that the absence of explicit electrostatics causes no significant problems, likely due to strong screening of electrostatic interactions. However, the absence of long-range electrostatics can cause anomalous water ordering in the $z$-direction, especially for models with a small receptive field. This issue leads to a broadening of the total dipole distribution, and thus likely affects work functions calculated from MLIP-based trajectories. Moreover, it should again be noted that the absence of explicit electrostatics can lead to issues when simulating electrolytes beyond a narrow interfacial region \citep{zhang2024molecular, kim2024learning}. Additionally, the effect of local fluctuations of the surface charge (i.e. image charges) in larger cells still requires further investigation. 

In the context of electrocatalysis, it is desirable for a single model to be able to handle different surface charges (i.e., different electrode potentials). Our results suggest that including electrostatics explicitly by learning the charges from local descriptors might not be sufficient to achieve this goal. Instead, we propose that MLIPs may benefit from having direct access to the global surface charge. A step in this direction is taken by constant-potential MLIPs \citep{chen2025grand, wang2025constant, chen2025constant, bergmann2025machine, chen2023atomistic}, which predict the Fermi level based on the number of excess electrons, which can be directly related to the surface charge. These models can be used to perform constant-potential molecular dynamics simulations, and do not require changing the electrolyte composition to simulate different surface charges. However, these models also require the use of implicit charging via continuum models which come with their own set of disadvantages as compared to explicit charging \citep{govindarajan2025intricacies}. Furthermore, given that the current implementations of such constant-potential MLIPs are still in early stages of development, further effort is required to provide accessible, well-documented, and efficient open-source implementation that can be adopted and validated by the wider community. 

Our results also demonstrate that pre-trained models such as eSEN-OC25 remain valuable for exploring configuration space and pre-equilibration when constructing new training sets, as well as for obtaining initial estimates of interfacial structures and reaction barriers. In addition, we found that MACE can run molecular dynamics simulations quite reliably even when trained on as few as 50 structures. While larger training datasets may be required for more complex systems involving ions and adsorbates, these results indicate that message-passing MLIPs with high-body-order descriptors can be used to explore the behavior of electrocatalytic interfaces with relatively little effort. 

\begin{acknowledgments}
This work originated from discussions at the 2024 Lorentz center workshop on Multiscale modeling of electrochemical processes (organized by Katharina Doblhoff-Dier, Nitish Govindarajan and Georg Kastlunger). N. G. acknowledges support from a startup grant at NTU (award number: 024462-00001). The authors thank the Lorentz center for the organization of the workshop. This publication is also part of the project “Computational Electrochemistry” with file number 2025.014 of the research programme “Computing Time on National Computer Facilities” which is (partly) financed by the Dutch Research Council (NWO) under the grant \url{https://doi.org/10.61686/BOKDD81349}.
\end{acknowledgments}

\onecolumngrid

\renewcommand{\thefigure}{S\arabic{figure}}
\renewcommand{\thesection}{S-\Roman{section}}
\renewcommand{\theequation}{S\arabic{equation}}
\renewcommand{\thetable}{S-\Roman{table}}
\setcounter{section}{0}
\setcounter{table}{0}
\setcounter{figure}{0}
\setcounter{equation}{0}

\section{The rotation group} \label{app:rotation}
Here, we provide a brief summary of the theory of the rotation group. For further reading, we refer the reader to the book of Zee \citep{Zee2016GTNutshell} for an introduction to group theory, Ref.~\onlinecite{duval2023hitchhiker} for an introduction to equivariant graph neural networks, and the books of Yutsis, Levinson and Vanagas \citep{Yutsis1962} and Marchildon \citep{Marchildon2002QuantumMechanics} for the theory of angular momentum in quantum mechanics.

Loosely speaking, a group is a collection of transformations that can be represented with a matrix, such as rotations, reflections or translations. The group of three-dimensional rotations and reflections is called O(3) (O for \emph{orthogonal}, because the group can be represented with orthogonal matrices). The group of only rotations is called SO(3) (S for \emph{special}). The group of translations, rotations and reflections is called the Euclidean group E(3).

\subsection{Rotating tensors}
In atomistic machine learning, one learns the relationship between a collection of points in three-dimensional space and some property (energy, forces, dipoles...). In this process, information is assigned to this collection of points. This information can be be stored in the form of scalars, vectors, or higher-rank tensors. When the system is rotated, this information needs to rotate accordingly (equivariance). The type of information that is stored can be characterized by how it transforms under rotation. For example, the rotation of a vector is described by a $3 \times 3$ rotation matrix $R$: $\vec{v} \mapsto R\vec{v}$. The matrices $R$ are $3 \times 3$ representations of SO(3). Scalars are invariant to rotations, so they transform as $s \mapsto s$. This identity transformation is the ($1 \times 1$) \emph{trivial} representation of SO(3).  A rank two tensor $T$ transforms with two rotation matrices, as 
\begin{equation}
    T_{ij} \mapsto \sum_{n=1}^3\sum_{m=1}^3 R_{in} R_{jm} T_{nm}.
\end{equation}
Note that a tensor is not just any kind of matrix: it is an object whose components transform into linear combinations of each other in a well-defined way. 

\subsection{Irreducible representations}
When we stack the 9 components of $T$ into a column vector, the transformation can be written in terms of a $9\times9$ rotation matrix $D(R)$. Such a rotation matrix is a $9 \times 9$ representation of SO(3). However, group theory states that this representation is \emph{reducible}: there are subsets of the 9 components that only transform into one another within that subset. In other words, $D(R)$ can be transformed by some matrix $S$ into a block-diagonal form
\begin{align}
    S D(R) S^{-1} &=\begin{pmatrix}  
    D^{(0)}(R) &&\\  
    & D^{(1)}(R) \\  
    && D^{(2)}(R)   
    \end{pmatrix} \\
    &= D^{(0)}(R) \oplus D^{(1)}(R) \oplus D^{(2)}(R) 
\end{align}
where $D^{(l)}(R)$ are the \emph{irreducible} representations (\emph{irreps}) of SO(3), and the entries left empty are zero. The irreducible representations are enumerated with a \emph{degree} $l$, and are matrices of size $(2l+1)\times(2l+1)$.

Hence, it turns out that some of the information in the tensor $T$ transforms like a scalar ($l=0$) and a vector ($l=1$). The only higher-degree information is encoded in the five components that transform according to the $l=2$ representation. As another example, the symmetric tensor $\vec{r}\otimes\vec{r}$ with six independent elements $r_ir_j$ can be transformed into one scalar element and five $l=2$ elements. In general, the irreps thus provide a more efficient way to store orientational information compared to Cartesian tensors. 

The $l$'th irrep space is spanned by $2l+1$ basis functions $\ket{lm}$, with $m$ an integer ranging from $-l$ to $l$. Vectors in the irrep spaces are referred to as spherical tensors. An $l=2$ spherical tensor can thus be written as $\vec A^{(2)}=A^{(2)}_2\ket{2 2} + A^{(2)}_1\ket{2 1} ... + A^{(2)}_{-2}\ket{2 (-2)}$. In position space, the kets $\ket{lm}$ are spherical harmonics: 
\begin{equation}
    \braket{\vec{x}|lm}=Y^{(l)}_m(\theta,\phi).
\end{equation}
The spherical harmonics form a basis for functions on the unit sphere; they do not depend on the radial coordinate $r$. 

\subsection{Clebsch-Gordan contraction} \label{app:cg-contraction}
In Cartesian space, two vectors $\vec{v}$ and $\vec{w}$ can be multiplied in an outer product to yield a rank two tensor: $T=\vec v\otimes \vec w$, i.e., $T_{ij}=v_iw_j$. We can do something similar with kets $\ket{lm}$. For example, in quantum mechanics, the state $\ket{l_1 m_1}\otimes\ket{l_2m_2}=\ket{l_1m_1;l_2m_2}$ can represent the combined state of two particles, where $l_1,m_1$ and $l_2,m_2$ are quantum numbers describing their angular momenta. The combined state rotates according to the rotation matrix $D^{(l_1)}(R)\otimes D^{(l_2)}(R)$. However, this representation is, in general, reducible:
\begin{equation} \label{eq:tensor-product-decomposition}
    C_{l_1l_2}^{-1} (D^{(l_1)}\otimes D^{(l_2)}) C_{l_1l_2} = D^{(|l_1-l_2|)} \oplus \cdots \oplus  D^{(l_1+l_2)}.
\end{equation}
The matrices $C_{l_1l_2}$ decompose the states $\ket{l_1m_1;l_2m_2}$ into irrep states $\ket{LM}$ that transform according to the corresponding irreducible representations on the right-hand side of Eq.~\ref{eq:tensor-product-decomposition}. This decomposition can be written as 
\begin{equation}
    \ket{l_1m_1;l_2m_2}=\sum_{L,M} C^{LM}_{l_1m_1l_2m_2} \ket{LM}
\end{equation}
where $$C^{LM}_{l_1m_1l_2m_2}=\braket{LM|l_1m_1;l_2m_2}$$ are elements of the matrices $C_{l_1l_2}$; they are the \emph{Clebsch-Gordan coefficients}. The Clebsch-Gordan coefficients are only nonzero for $M=m_1+m_2$ and for integers $L$ from $|l_1-l_2|$ to $l_1+l_2$. Sometimes, the states $\ket{LM}$ are also written $\ket{l_1l_2LM}$.

In quantum mechanics, $L$ and $M$ describe the angular momentum of the combined system. The state $\ket{l_1m_1;l_2m_2}$ thus does not have a well-defined total angular momentum $L$, whereas $\ket{LM}$ \emph{does}. To `build' states with a well-defined $L$ in terms of states $\ket{l_1m_1;l_2m_2}$, we can write 
\begin{equation}
    \ket{LM} = \sum_{m_1m_2} C^{LM}_{l_1m_1l_2m_2} \ket{l_1m_1;l_2m_2}.
\end{equation}
This transformation, in which combined states with $(2l_1+1)(2l_2+1)$ components are reduced into irrep states of $(2L+1)$ components, is also referred to as Clebsch-Gordan \emph{contraction}.

For higher-order tensor products of angular momentum states, such as 
\begin{equation}
\ket{l_1m_1}\otimes\ket{l_2m_2}\otimes\ket{l_3m_3}\otimes\cdots,
\end{equation}
the contraction is described by generalized Clebsch-Gordan coefficients $\mathcal C_\mathbf{lm}^{LM}$ with multi-indices $\mathbf{l}=(l_1,l_2...)$ and $\mathbf{m}=(m_1,m_2,...)$\citep{Yutsis1962}.

\section{Training details} \label{si:training_details}
In this work we first trained models to test their data efficiency when trained on small training sets, and their accuracy when trained on the full training sets. Whereas the DP and DP-MP implementation defines training duration by the number of gradient updates, GRACE and MACE define training duration by the number of epochs. Because of the different training set sizes used in the data efficiency tests, it is important to train models with a similar number of gradient updates (GU), where
$$
\text{number of GU} = \text{number of epochs} \times \text{dataset size} / \text{batch size}.
$$
This consideration led to the hyperparameters listed in Table \ref{tab:training_data_efficiency}. When training GRACE (both 1-layer and 2-layer) and MACE on small datasets, the loss already plateaued far before reaching 100 000 gradient updates. For this reason and to reduce computational cost, MACE models were not trained for more than 2000 epochs and GRACE models not more than 4000 epochs.

\begin{table}[htbp]
\centering
\caption{Training hyperparameters for the data efficiency tests. For GRACE and MACE, the number of epochs is indicated separately for training datasets of 50, 100, 500 and 1000 structures. $w_{E,F}$ are the weights of energy and force errors in the loss function.}
\label{tab:training_data_efficiency}
\begin{tabular}{llcl}
\hline
\textbf{Model} & \textbf{Training length} & \textbf{Batch size} & \textbf{Optimization} \\
\hline
DP, DP-MP & 150k gradient updates & 3 & LR=0.002, exponential decay \\
GRACE & 4000, 4000, 800, 400 epochs & 3 & LR=0.008, $w_E=1$, $w_F=5$, cosine decay \\
MACE & 2000, 2000, 800, 400 epochs & 3 & LR=0.01, $w_E=1$, $w_F=10$, \\
&&& after 60\% of total epochs: \\
&&& LR=0.001, $w_E=1000$, $w_F=100$ \\
\hline
\end{tabular}
\end{table}

\begin{figure}[h]
    \centering
    \includegraphics[width=4in]{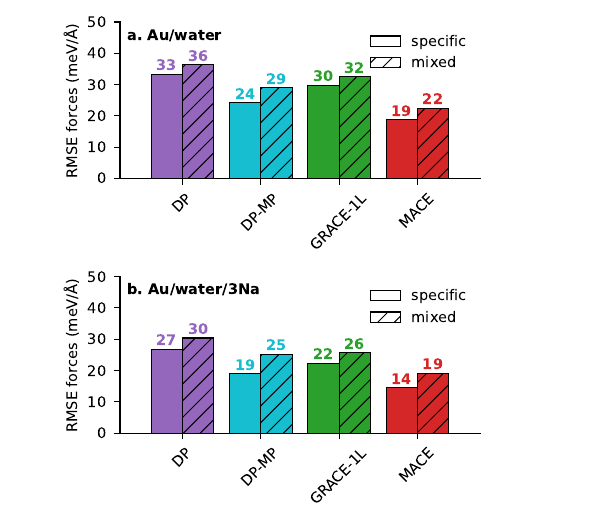}
    \caption{(a) Force RMSE on the neutral-surface Au/water test set with 2690 structures for models trained on the target system (`specific') vs. models trained on the mixed dataset. (b) Force RMSE on the negative-surface Au/water/3\ce{Na^+} test set with 2463 structures for specific models vs. mixed models. }
    \label{fig:ion-water-force-rmse}
\end{figure}

For the comparison of specific and mixed models, training was extended slightly to ensure full convergence. An overview of the training hyperparameters is given in Table \ref{tab:training_full}. For the DP and DP-MP models, a larger batch size made the loss decay more monotonously, but a smaller batch size slightly improved the accuracy for the models trained on data with ions. Training longer did not affect the trend in the test force RMSE, as can be seen by comparing the errors on the largest training set in Fig.~2 in the main text with Fig.~\ref{fig:ion-water-force-rmse}a. The force errors on the negative-surface system show the same trend as well (Fig.~\ref{fig:ion-water-force-rmse}b). However, the energy errors shown in Figures~4 and~6 in the main text show slightly different behavior.

\begin{table}[htbp]
\centering
\caption{Training hyperparameters for full dataset training.}
\label{tab:training_full}
\begin{tabular}{llcl}
\hline
\textbf{Model} 
& \textbf{Training length} 
& \textbf{Batch size} 
& \textbf{Optimization} \\
\hline

DP, DP-MP 
& 300k gradient updates
& 3 
&  LR=0.002, exponential decay \\
(no ions) \\

DP, DP-MP 
& 1 million gradient updates 
& 1 
&  LR=0.002, exponential decay  \\
(with ions) \\

GRACE-1L 
& 500 epochs 
& 3 
& LR=0.01, $w_E=1$, $w_F=5$. \\
&&& After 300 epochs: LR=0.001, $w_E=30$, $w_F=1$ \\

MACE 
& 500 epochs 
& 3 
& LR=0.01, $w_E=1$, $w_F=10$. \\
&&& After 300 epochs: LR=0.001, $w_E=1000$, $w_F=100$ \\

\hline
\end{tabular}
\end{table}

Furthermore, 32-bit float precision was used for DP, DP-MP and MACE, and 64-bit float precision for GRACE.

\section{Active learning} \label{si:active-learning}
An initial dataset created using a MACE active learning routine was obtained from Ref.~\onlinecite{ankit2025inpreparation}. This dataset contained roughly 500 structures for each type of the aforementioned systems. 

DP and DP-MP models were trained on the initial dataset, but were unstable initially, i.e., they showed large changes in the total energy during MD simulations. Often, this was due to ions approaching the surface too closely. From these trajectories, structures were sampled in 2 ways: (1) random sampling before the Nosé-Hoover invariant energy starts deviating more than a threshold 0.1\%; (2) sampling frames from the last \SI{0.1}{\pico\second} before the invariant energy starts deviating more than the threshold. 

The structures sampled from these trajectories were added to the existing dataset. The Maximum Set Coverage (MSC) algorithm from Ref.~\onlinecite{yu2025maximizing}, implemented in the \texttt{QUESTS} package \citep{schwalbekoda2025information}, was employed to select the $N$ most diverse configurations from the total data pool, with $N$ being the targeted size of the training set. The structures in this selection were labeled with DFT. 

After labeling, structures with energies deviating more than \SI{30}{\electronvolt} (corresponding to roughly \SI{80}{\milli\electronvolt\per atom}) from the mean energy of that type of structure, or with forces larger than \SI{15}{\milli\electronvolt\per\angstrom}, were discarded. Discarding these unlikely configurations ensures that the models are trained on configurations similar to those that are typically encountered in molecular dynamics simulations at \SI{300}{\kelvin}.

With the new dataset including the additional configurations, the DP and DP-MP models were retrained, and several molecular dynamics runs were used to check for stability. The above procedure was repeated until the invariant energy was stable over \SI{2}{\nano\second}.

Only the DP and DP-MP models were used for active learning because these models appeared the most unstable and are computationally cheap. However, all models were trained on the same final datasets for analysis.


\section{Uncertainty estimate for and reweighting of density histograms} \label{si:reweighting}
The uncertainty associated with MLIP training in the ion density histograms was determined from the spread predicted in the histogram from different committee members. To reduce computational cost, the ion densities of different committee members was not computed directly from individual molecular dynamics runs. Instead histogram reweighting was used and only the first model committee member was used to run a \SI{2}{\nano\second} trajectory. 

In principle reweighting can be done exactly by recalculating the histogram with frame weights
\begin{equation}
    w(q)=\exp(-\beta[E^{(i)}(q) - E^*(q)]),
\end{equation}
where $q$ is a trajectory frame, $\beta=1/k_\mathrm{B}T$, $E^{(i)}(q)$ the energy predicted by committee member $i$ and $E^*(q)$ the energy from the potential that was used to originally run the trajectory. However, Ref.~\onlinecite{imbalzano2021uncertainty} suggests using the statistically more stable cumulant expansion approximation (CEA). The histogram according to committee member $i$ is estimated using the CEA as
\begin{equation}
\langle n_b \rangle_{E^{(i)}} \approx\; \langle n_b \rangle_{E^*} - \beta\Big[\langle n_b (E^{(i)} - E^*) \rangle_{E^*} - \langle n_b \rangle_{E^*} \langle E^{(i)} - E^* \rangle_{E^*}\Big]
\end{equation}
where $n_b$ is the count of ions in histogram bin $b$, and $\langle ...\rangle_{E^*}$ implies averaging over the trajectory driven by $E^*$. The mean and standard deviation was calculated over the histograms from $M=5$ committee member predictions, and the standard deviation was scaled as $\sigma \to \alpha\sigma$.

The scaling factor $\alpha$ is introduced to correct for an underestimation of the committee uncertainty. The unbiased estimator \citep{imbalzano2021uncertainty}
\begin{equation}
    \alpha^2 = -\frac{1}{M} + \frac{M-3}{M-1} \frac{1}{N_\mathrm{val}} \sum_{q\in\mathrm{val}} \frac{(E_\mathrm{ref}(q)-\bar{E}(q))^2}{s^2(q)}
\end{equation}
was used, where $M$ is the number of models trained, $N_\mathrm{val}$ is the number of configurations in the validation set, $\bar{E}(q)$ is the mean of committee energy predictions for the validation set, $E_\mathrm{ref}(q)$ the true energy, and $s^2(q)$ the committee sample variance for the energy prediction of structure $q$ in the validation set (val). The $\alpha$ values calculated for the different models are shown in Table \ref{tab:alpha_values}. Generally, the mixed models have a higher $\alpha$ than the specific models, which contributes to their high uncertainty.

\begin{table}
\centering
\caption{Estimated calibration constant $\alpha$ for different models.}
\label{tab:alpha_values}
\begin{tabular}{lcc}
\hline
model & three-ion & mixed \\
\hline
DP    & 2.1 & 5.4 \\
DP-MP  & 1.0 & 8.1 \\
GRACE-1L & 7.7 & 12.0 \\
MACE-e  & 4.1 & 6.7 \\
\hline
\end{tabular}
\end{table}

\begin{figure}[h]
    \centering
    \includegraphics[width=5.5in]{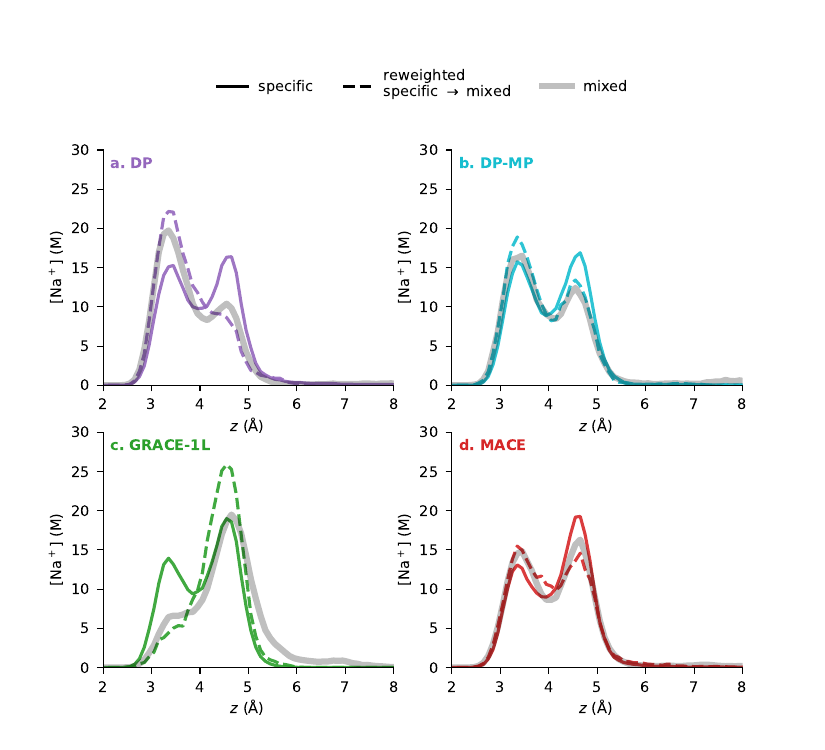}
    \caption{Reweighting density histograms. For each architecture, the density histogram obtained from a trajectory of the specific model is reweighted with the mixed potential. The true mixed trajectory density histograms are also shown in light gray for comparison.}
    \label{fig:check-reweighting}
\end{figure}

As an example of reweighting histograms with the CEA, we show in Fig.~\ref{fig:check-reweighting} that the reweighting technique can be used to study the effect of another potential. The density histogram obtained from a \SI{2}{\nano\second} trajectory (solid colored lines) from MD driven by a specific-charge model is reweighted with a mixed potential of the same model architecture (dashed lines). The reweighted histograms give a similar result to the true histogram obtained from a simulation driven by the mixed-data model (thick gray lines), although the peak height is exaggerated in some cases. Some deviation is to be expected due to finite sampling and the cumulant expansion formula only being an approximation. 

\clearpage

\section{Water dynamics} \label{si:water-dynamics}
We calculated the water dipole autocorrelation function
\begin{equation}
    C_\mathrm{dip}(\tau) = \langle \vec{p}_i(t)\cdot\vec{p}_i(t+\tau)\rangle_{t,i\in\mathrm{interface}}
\end{equation}
where $\vec{p}_i$ are the normalized water bisectors for the water molecules $i$ in the interfacial region (0 to \SI{4.5}{\angstrom}). The range of $\tau$ used was $\tau=0$ to \SI{2.5}{\pico\second} and the averages were taken over separate trajectory blocks ($t=0$ to \SI{2.5}{\pico\second}, \SI{2.5}{\pico\second} to \SI{5}{\pico\second}, etc.).

We also calculated the survival probability of a water molecule in the interfacial region as
\begin{equation}
    P_\mathrm{surv}(\tau)= \langle \mathbb{I}_i(t)\mathbb{I}_i(t+\tau)\rangle_{t,i\in\mathrm{interface}}
\end{equation}
where $\mathbb{I}_i(t)$ is the indicator function, indicating whether particle $i$ is in the interfacial region (0 to \SI{4.5}{\angstrom}) at time $t$. $P_\mathrm{surv}$ was calculated from $\tau=0$ to \SI{5}{\pico\second}, again averaged over separate trajectory blocks. The code is available at \url{github.com/lucasdekam/WatAnalysis} \citep{dekam_watanalysis2025}.

\begin{figure}[h!]
    \centering
    \includegraphics{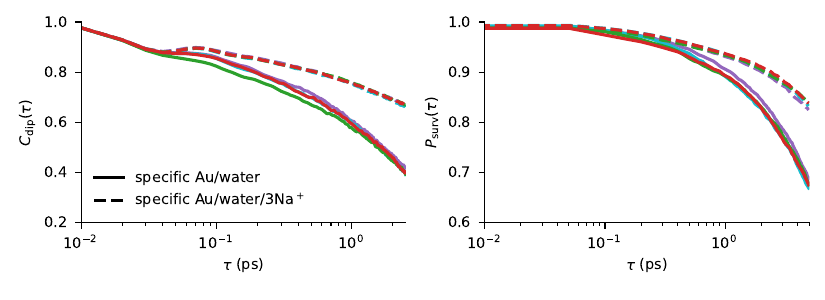}
    \caption{Dipole autocorrelation function (left) and water survival probability (right) for interfacial water (0 to \SI{4.5}{\angstrom} from the surface) in the Au/water and Au/water/3\ce{Na^+} systems, using DP (purple), DP-MP (cyan), GRACE-1L (green), MACE (red).}
    \label{fig:water-dynamics-all}
\end{figure}

From the dipole autocorrelation function and water survival probability in Fig. \ref{fig:water-dynamics-all}, we see that the outputs from molecular dynamics simulations with different model architectures closely agree. The dipole autocorrelation function from the GRACE-1L trajectory seems to deviate slightly. The water at the charged Au/water/3\ce{Na^+} interface is more rigid: the water dipoles are more strongly correlated and water has a higher probability of staying in the interfacial region.

\begin{figure}[h!]
    \centering
    \includegraphics{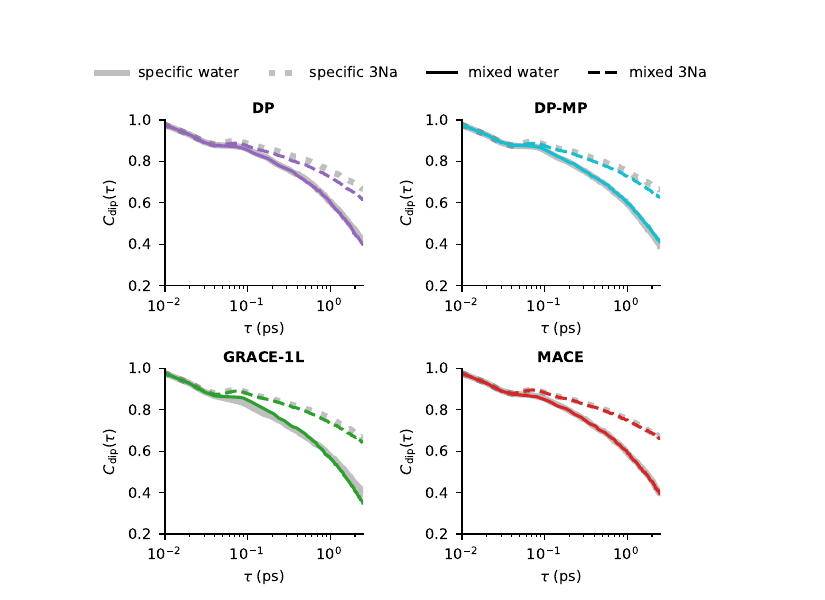}
    \caption{Comparison of dipole autocorrelation function in the interfacial region (0 to \SI{4.5}{\angstrom}) from simulations with specific and mixed-data models for Au/water (solid lines) and Au/water/3\ce{Na^+} interfaces (dashed). The results for the specific models (same results as in Fig. \ref{fig:water-dynamics-all}) are shown in light gray.}
    \label{fig:water-dynamics-dip-corr}
\end{figure}

Figure \ref{fig:water-dynamics-dip-corr} compares dipole autocorrelation functions between specific and mixed models. The mixed and specific models agree nearly perfectly for the Au/water interfaces, showing little signs of data interference. For the Au/water/3\ce{Na^+} interface, the MACE mixed model result is indistinguishable from the specific model result. For the other architectures, the mixed model predicts slightly less correlated water compared to the specific model, likely a sign of data interference.

\clearpage

\section{Supplementary figures}


\begin{figure}[h]
    \centering
    \includegraphics[width=5.5in]{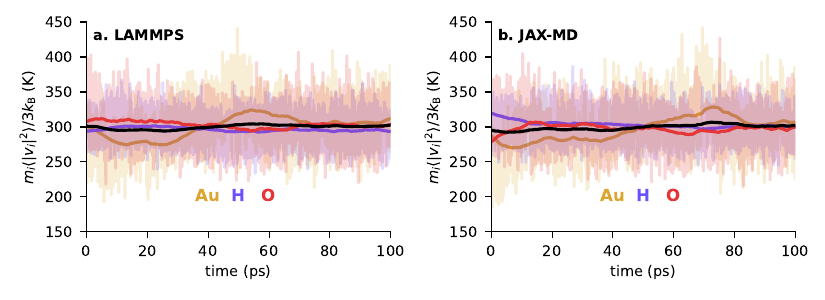}
    \caption{Average kinetic energy in K for the different species in a simulation of a gold-water interface driven by (a) a MACE potential in LAMMPS and (b) a DP potential in JAX-MD, both with a Nosé-Hoover chain thermostat. The full energy fluctuations are shown in light colors, the darker-colored lines show an exponential moving average. The black line shows the exponential moving average of the mean of all species.}
    \label{fig:fig-thermostats}
\end{figure}

\begin{figure}[h]
    \centering
    \includegraphics[width=5.5in]{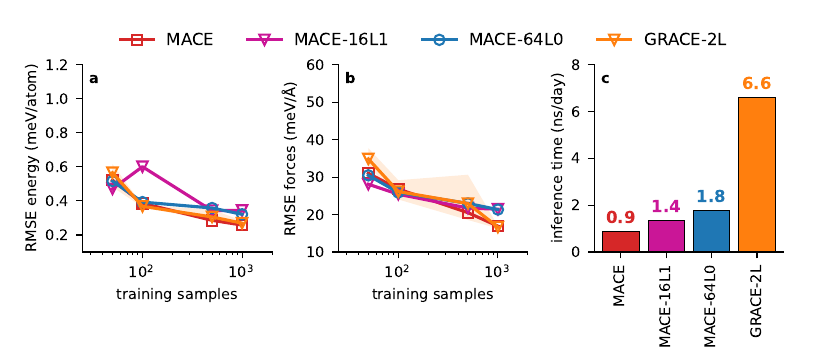}
    \caption{Accuracy and performance of three variants of MACE, compared to 2-layer GRACE (GRACE-2L). MACE-16L1 has 16 scalar and 16 vector node features ($L_\mathrm{max}=1$). MACE-64L0 has 64 scalar node features ($L_\mathrm{max}=0$). All MACE models have two layers. GRACE-2L uses the default setup of the \texttt{gracemaker} package with $c=\SI{5}{\angstrom}$; it is equivariant with $L_\mathrm{max}=1$. (a-b): root-mean-squared error (RMSE) of the energy (a) and force components (b) on a test set containing 200 neutral Au/water interface structures. (c) Molecular dynamics speed for an Au/water interface with 384 atoms and a \SI{0.5}{\femto\second} timestep on an NVIDIA A100 GPU.  }
    \label{fig:rmse-cost-si}
\end{figure}

\clearpage

\begin{figure}[h]
    \centering
    \includegraphics[width=5.5in]{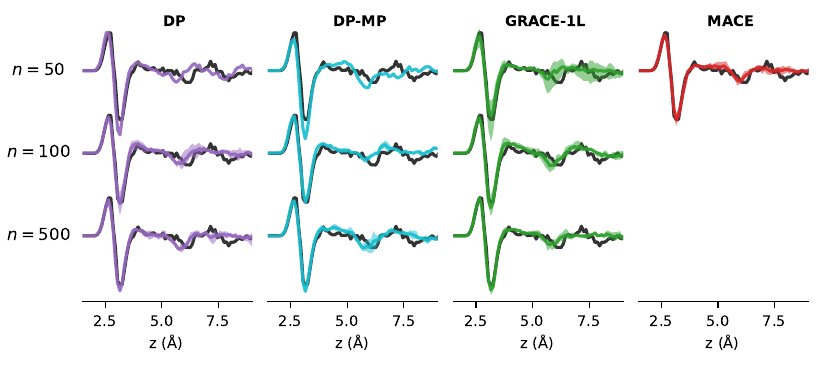}
    \caption{Water orientation profiles in the direction perpendicular to the Au surface obtained from \SI{100}{\pico\second} trajectories driven by models trained on different amounts of training data, for the neutral-surface dataset. For each model and training set, the mean of three repeats is shown (in color) with models trained on different subsets of the data and a different initial configuration. The spread indicates the min/max deviation over the three reruns. Black lines indicate the mean result from three 10-ps DFT-MD trajectories.}
    \label{fig:water-orientation-si}
\end{figure}

\begin{figure}[h!]
    \centering
    \includegraphics[width=5.5in]{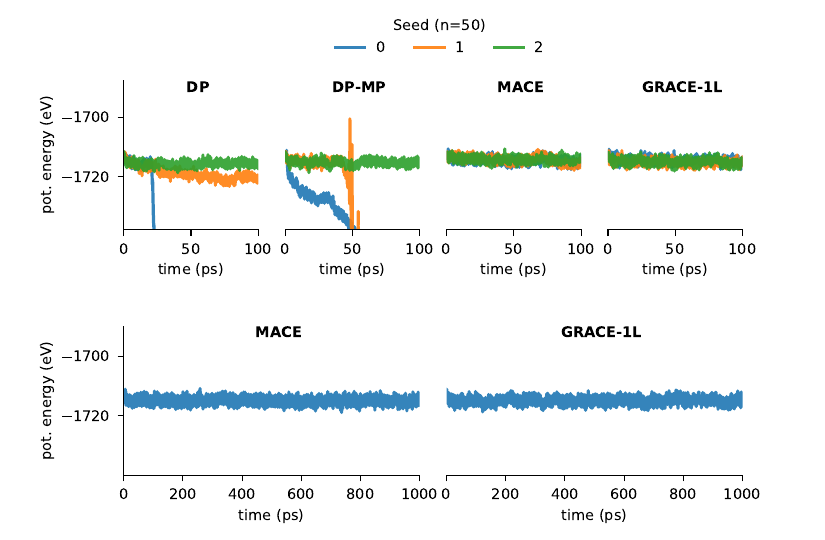}
    \caption{Potential energy over the course of trajectories of the Au/water system for models trained on three different samples of $n=50$ structures. For seed 0, MACE and GRACE-1L were also run for \SI{1}{\nano\second}, as shown in the bottom half of the figure.}
    \label{fig:trajectory-energy}
\end{figure}


\begin{figure}[h]
    \centering
    \includegraphics[width=5.5in]{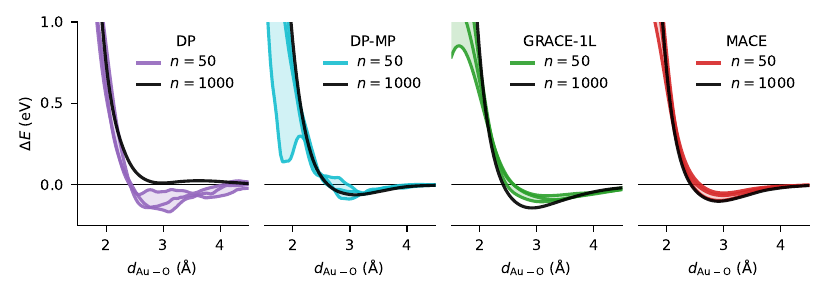}
    \caption{The learned potential energy surface for a single O-down water molecule approaching an Au surface. The colored lines indicate curves evaluated with the models trained on 50 samples (three for each model type); the black lines indicate curves evaluated with the final models trained on 1000 samples. Note that the PESes might differ between models due to the way they decompose the many-body energy when training on Au surfaces with many water molecules; this figure shows the roughness of the PES, not the correctness.}
    \label{fig:pes-cut}
\end{figure}

\begin{figure}[h]
    \centering
    \includegraphics[width=3.5in]{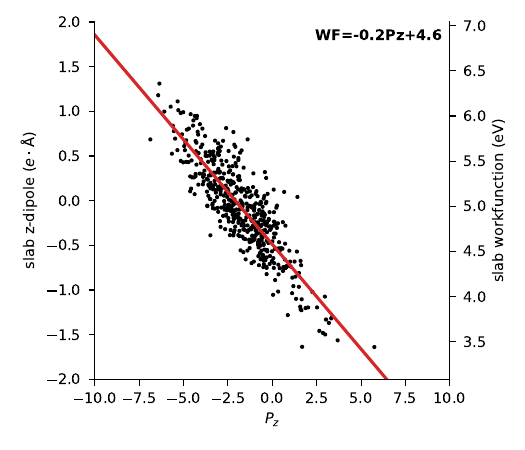}
    \caption{Slab z-dipole and work function obtained from DFT for frames taken from the ab initio trajectories of the gold-water system, compared to the total geometric dipole $P_z$. An increase of 5 units on the $P_z$ axis roughly corresponds to a decrease in work function of 1 eV. }
    \label{fig:dipoles}
\end{figure}

\begin{figure}[h]
    \centering
    \includegraphics[width=5.5in]{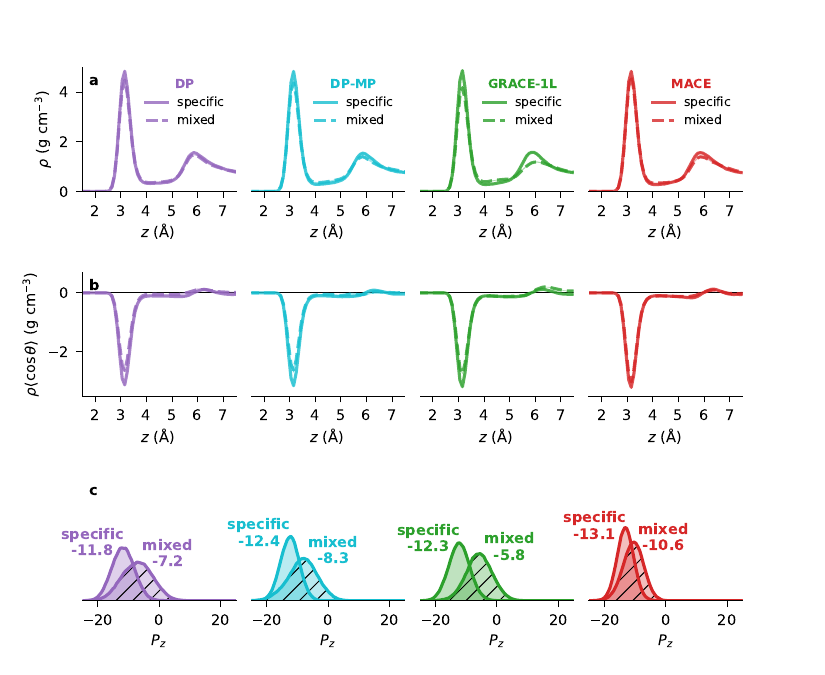}
    \caption{Water structure in the Au/water/3\ce{Na^+} system, comparing models trained only on Au/water/3\ce{Na^+} data (specific) versus models trained on a dataset containing interfaces with 0 to 4 \ce{Na^+} ions (mixed). (a) Water density and (b) dipole orientation profiles. (c) Total dipole ($P_z$) distributions. The mean values of the distributions are indicated in the plot.}
    \label{fig:ions}
\end{figure}

\clearpage

\section*{References}
\bibliography{aipsamp}

\end{document}